\documentclass[aps, reprint, prl, twocolumn, superscriptaddress,amsmath,amssymb,noeprint]{revtex4-2}
\usepackage{times}
\usepackage{amsmath}
\usepackage{amssymb}
\usepackage{xfrac}
\usepackage{dsfont}
\usepackage{graphicx}
\usepackage{braket}
\usepackage{bbold}
\usepackage{bm}
\usepackage{textcomp}
\usepackage{bbm}
\usepackage{color}
\usepackage{xcolor}
\usepackage[normalem]{ulem}
\usepackage{pdfpages}
\usepackage{pgffor}
\usepackage{float}
\usepackage{gensymb}
\usepackage{caption} 
\usepackage{ragged2e} 

\usepackage{mathtools}
\usepackage{nccmath}
\usepackage[british]{babel} 
\usepackage{sidecap}
\usepackage{soul}
\usepackage{upgreek}
\usepackage{enumitem}
\usepackage[linesnumbered, ruled]{algorithm2e}
\usepackage{verbatim}
\usepackage{etoolbox}
\usepackage{makeidx}
\usepackage{capt-of}
\usepackage{esvect}

\usepackage{tocloft}

\DeclareCaptionJustification{capjust}{\justifying}

\makeatletter
\AtBeginDocument{\let\LS@rot\@undefined}

\renewcommand{\section}{\@startsection
  {section}{1}{1.8pt}
  {4.2ex plus 0.5ex minus 0.2ex}  
  {1.4ex}                         
  {\normalfont\bfseries\centering}
}

\makeatother

\usepackage[colorlinks=true, urlcolor=blue, linkcolor=blue, citecolor=blue]{hyperref}
\usepackage{siunitx}

\newcommand{\berkeleyphy}{Department of Physics, University of California, Berkeley, California 94720}
\newcommand{\CIQC}{Challenge Institute for Quantum Computation, University of California, Berkeley, California 94720}
\newcommand{\LBL}{Materials Sciences Division, Lawrence Berkeley National Laboratory, Berkeley, California 94720}
\newcommand{\JILA}{JILA, National Institute of Standards and Technology and University of Colorado,
Boulder, Colorado 80309}

\begin{document}

\title{Astigmatism-free 3D Optical Tweezer Control for Rapid Atom Rearrangement}

\author{Yue-Hui Lu}
\affiliation{\berkeleyphy}
\affiliation{\CIQC}

\author{Nathan Song}
\affiliation{\berkeleyphy}
\affiliation{\CIQC}
\affiliation{\JILA}

\author{Tai Xiang}
\affiliation{\berkeleyphy}
\affiliation{\CIQC}

\author{Jacquelyn Ho}
\affiliation{\berkeleyphy}
\affiliation{\CIQC}

\author{Tsai-Chen Lee}
\affiliation{\berkeleyphy}
\affiliation{\CIQC}

\author{Zhenjie Yan}
\affiliation{\berkeleyphy}
\affiliation{\CIQC}

\author{Dan M. Stamper-Kurn}
\email[]{dmsk@berkeley.edu}
\affiliation{\berkeleyphy}
\affiliation{\CIQC}
\affiliation{\LBL}


\begin{abstract} 
Reconfigurable neutral-atom arrays are a promising platform for quantum computing, quantum simulation, and quantum metrology, but atom transport using frequency-chirped acousto-optic deflectors (AODs) is limited by chirp-induced acoustic lensing and trajectory distortion. We address these limitations using a three-dimensional acousto-optic deflector lens (3D-AODL), a design predicted to reduce long-range transport times by more than a factor of two. We further introduce fading-Shepard waveforms that circumvent finite AOD bandwidth, enabling sustained axial displacement. We demonstrate unrestricted three-dimensional optical-tweezer motion over a 200 $\mu$m × 200 $\mu$m × 136 $\mu$m volume with velocities exceeding 4.2 m/s. Arbitrary three-dimensional control of optical-tweezer trajectories enables rapid atom rearrangement and dynamical engineering of optical potentials in tweezer arrays and optical lattices. This capability advances quantum control and atom manipulation in neutral-atom quantum processors by enabling faster rearrangement, higher clock rates, and scalable sorting in complex geometries.
\end{abstract}

\maketitle

\begingroup
\let\oldaddcontentsline\addcontentsline
\renewcommand{\addcontentsline}[3]{}

\section{Introduction}
Optical tweezers capable  of strong, highly tunable spatial confinement of microscopic objects are used to trap atoms~\cite{Browaeys2020}, molecules~\cite{Kaufman2021}, nanoparticles~\cite{MacDonald2005,Ott2022}, cells~\cite{Arbore2019}, and DNA~\cite{Smith2003}. Confinement occurs because the intense electric field gradient of a focused laser induces an optical dipole force on particles that pulls them to the point of highest intensity. 
Due to their highly programmable nature, optical tweezer arrays have become a leading platform for neutral atom quantum computing~\cite{ Jaksch2000, Wilk2010, Graham2022, Bluvstein2024, Henriet2020} and quantum simulation~\cite{Ebadi2021, Morgado2021}.

A variety of optical devices have been used to generate optical tweezers. For applications requiring dynamical variation of optical tweezers at microsecond timescales, such as those involving the physical transport of tweezer-trapped atoms and molecules, acousto-optic deflectors (AODs) are preferred for tweezer generation over slower devices such as liquid-crystal spatial light modulators (SLMs)~\cite{Kim2019} or digital micromirror devices (DMDs)~\cite{Stuart2018}.  While MHz \cite{simon2026, vuletic2026} and GHz \cite{Peng19} SLM architectures have been explored, limitations in power efficiency and scalability have thus far kept these approaches largely at the developmental stage, while commercially available devices remain limited to kHz refresh rates that are too slow for rapid tweezer motion. 
As such, AODs remain the most practical option for dynamic tweezer control.  An AOD steers light by using a piezoelectric transducer to drive a radio frequency (rf) acoustic wave in an acousto-optic crystal to create a tunable diffraction grating~\cite{ Korpel1988Acousto}. Changing the driving frequency shifts the diffraction angle, allowing beam steering with MHz bandwidth.

An AOD operates with a running acoustic wave and thus its spatial and temporal responses are intrinsically linked. The finite speed of sound limits the acoustic transit time within the AOD crystal to $\sim\!10\,\mu$s. Because of this delayed response, changes in the rf drive induce aberrations in the tweezer profile~\cite{Korpel1981}. Notably, if the rf drive on a one-dimensional AOD is chirped linearly, the aberration is astigmatic, with the AOD now acting as a cylindrical lens with a dioptric power that scales with the chirp rate~\cite{Kaplan2001}. Such aberration has two negative effects on atom transport in optical tweezers: it deforms the optical tweezer trap profile, reducing its depth and stability, and it causes the locus of maximum intensity of the tweezer to trace an unwanted out-of-plane trajectory. 
Atom rearrangement time is primarily set by two processes: (i) the pick-up/drop-off of atoms in static tweezers, and (ii) their transport between sites. While pick-up/drop-off typically requires hundreds of microseconds, transport time scales with distance and introduces motional heating \cite{hwang2025fast}.  To minimize atom loss, experiments use relatively slow transport speeds, empirically found to be about 0.5 m/s~\cite{Manetsch2025, Bluvstein2022}.  
For long-range motion, this scaling can lead to substantially longer rearrangement times, making transport speed limit a key bottleneck for large-scale atom arrays and zone-based architectures \cite{gyger2024continuous, Bluvstein2024, Manetsch2025,  Chiu2025, Lin2025}, especially for fly-by gate protocols where atoms remain in motion and entangling operations are performed on-the-fly\cite{lib2026}.

\begin{figure*}
{\includegraphics[width=1.0\textwidth]{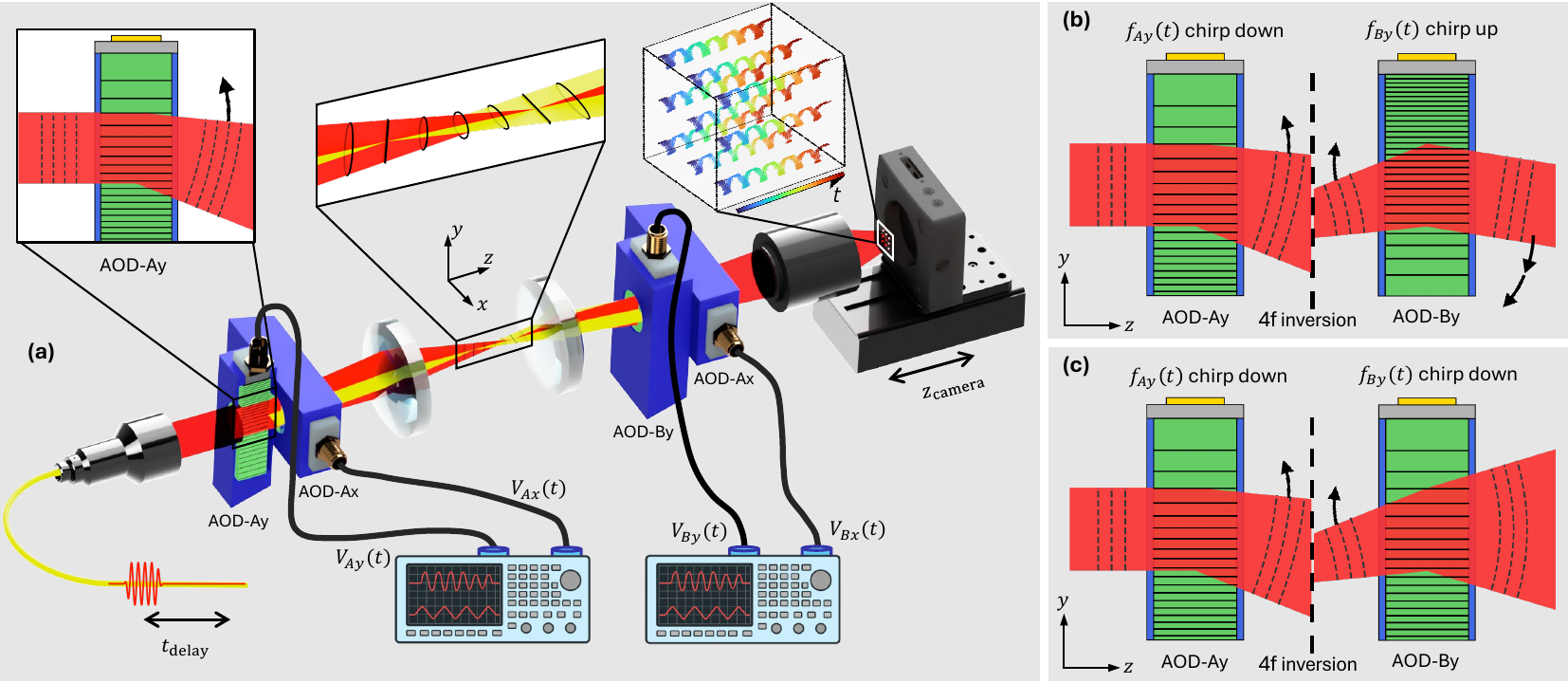}}
    \caption{
    \textbf{3D-AODL setup for time-resolved 3D light field tomography.} 
    (a) Experimental schematic. Chirped rf drives applied to the first 2D-AODs (Ax, Ay) create cylindrical lensing; the left inset shows acoustic (solid) and optical (dashed) wavefronts, resulting in astigmatism in the intermediate plane (middle inset); This aberration is compensated by the second 2D-AODs (Bx, $By$), which are configured in a $4f$ arrangement relative to the first. The output is Fourier-imaged onto a camera with a $F=100$\,mm objective. Scanning the camera position and light-pulse delay renders time-resolved tomography of the tweezer light field. The right inset shows a reconstructed light field, with color indicating the flow of time (red to blue). 
    (b) Beam steering without lensing, when parallel AODs are driven with counter-chirped frequencies whose sum is constant. 
    (c) Lensing without steering, when parallel AODs are driven with co-chirped frequencies whose difference is constant.
    Note: ``4f inversion'' refers to propagation through the 4f system in (a), omitted for brevity. 
    }

    \label{fig: M1}
\end{figure*} 

While several research efforts have focused on optimizing tweezer trajectories to eliminate mechanical excitation in and loss of atoms from the accelerration and jerk of  tweezers ~\cite{kim2025blinkingopticaltweezersatom, hwang2023optical, hwang2025fast, Cicali2025, pagano2024}, the underlying issue of chirp-induced aberration has been largely unaddressed. Previous work~\cite{Reddy2005, Kirkby2010} in optics demonstrated that a three-dimensional acousto-optic deflector lens (3D-AODL) composed of four AODs could control chirp-induced acousto-optic lensing both to minimize aberrations and to access the axial spatial direction. These methods were highly effective for bio-imaging~\cite{Szalay2016, Nadella2016, Katona2012} and microscopy applications~\cite{Soto:18}  that utilized high-speed scanning of 3D optical intensity profiles. However, a challenge for these systems is to produce sustained axial displacement: a continuous rf chirp within the fixed bandwidth of the AOD device can only be maintained for a finite time.  A simple remedy to the chirp-time limitation is to apply a serrodyne waveform, discontinuously interrupting a monotonic chirp that reaches one end of the AOD bandwidth and instantaneously restarting the rf chirp at the other end of the bandwidth~\cite{Kirkby2010}.  However, such saw-tooth frequency modulation would produce a periodic flicker in the tweezer intensity strength and spatial pattern, disturbing the particle trapped within.

In this work, we report a 3D-AODL device composed of four AODs in a 4f imaging configuration, as shown in Fig.~\ref{fig: M1}(a), which enables aberration-free, omnidirectional 3D tweezer motion. We introduce a novel family of \emph{fading-Shepard} waveforms that allows for sustained and flicker-free axial displacement. We perform time-resolved 3D tomography of the tweezer trajectories via characterizing the optical tweezer intensity field produced by this device using stroboscopic imaging techniques and an axial translation stage.
We predict significant improvements in the efficiency of atom transport. Monte Carlo simulations predict a 50\% - 75\% reduction in long-range transport time compared to conventional methods. This improvement could enable faster computational clock rates and reduced atom loss in neutral-atom-based quantum computers. Furthermore, fading-Shepard waveforms enables rapid ($\sim$10s of $\mu s$) and unconstrained access to the axial dimension of the tweezers, enabling advanced rearrangement and modulation in 3D atom arrays.
Spatial coordinates are reported in converted units, $\mu\text{m}^*$, to reflect typical atom trapping scales, as detailed in Sec.~2A in the Supplementary Material.

\section{Astigmatism-free tweezers}

When optimizing atom transport in an optical tweezer, it is commonly assumed that the shape of the tweezer trapping potential stays fixed in the moving frame. Various works have optimized tweezer trajectories under this assumption~\cite{hwang2025fast, Cicali2025, pagano2024}, which is valid for low-speed and short-range transport. However, for long-range rapid transport, the rigid tweezer picture fails due to astigmatic aberration that deforms the trap profile of the moving tweezer.  To illustrate this point, in Fig.~\ref{fig: M2} we present the calculated probability that an atom will remain trapped in an optical tweezer trap after it is rapidly displaced along two different trajectories: a rapid displacement produced either by chirping one or both AODs within a two-dimensional AOD setup.  We quantify this probability for minimum-jerk trajectories with fixed displacement ($65\,\mu\text{m}$ for the $X$-motion and $65\sqrt{2}\,\mu\text{m}$ for the diagonal motion) at various transport times, comparing results for tweezers that are aberrated by uncompensated AOD lensing to those for which the intensity profile of the translated tweezer remains constant during its translation (see Supplement, Sec. 1A.3, where we also compare the mininum-jerk ramp with other forms of ramps).  The survival probability for the aberration-uncompensated tweezers is significantly worsened, requiring much longer transport times to avoid atom loss and even longer times to avoid motion-induced heating.

\begin{figure*}[t]
    {\centering
    \includegraphics[width=1\linewidth]{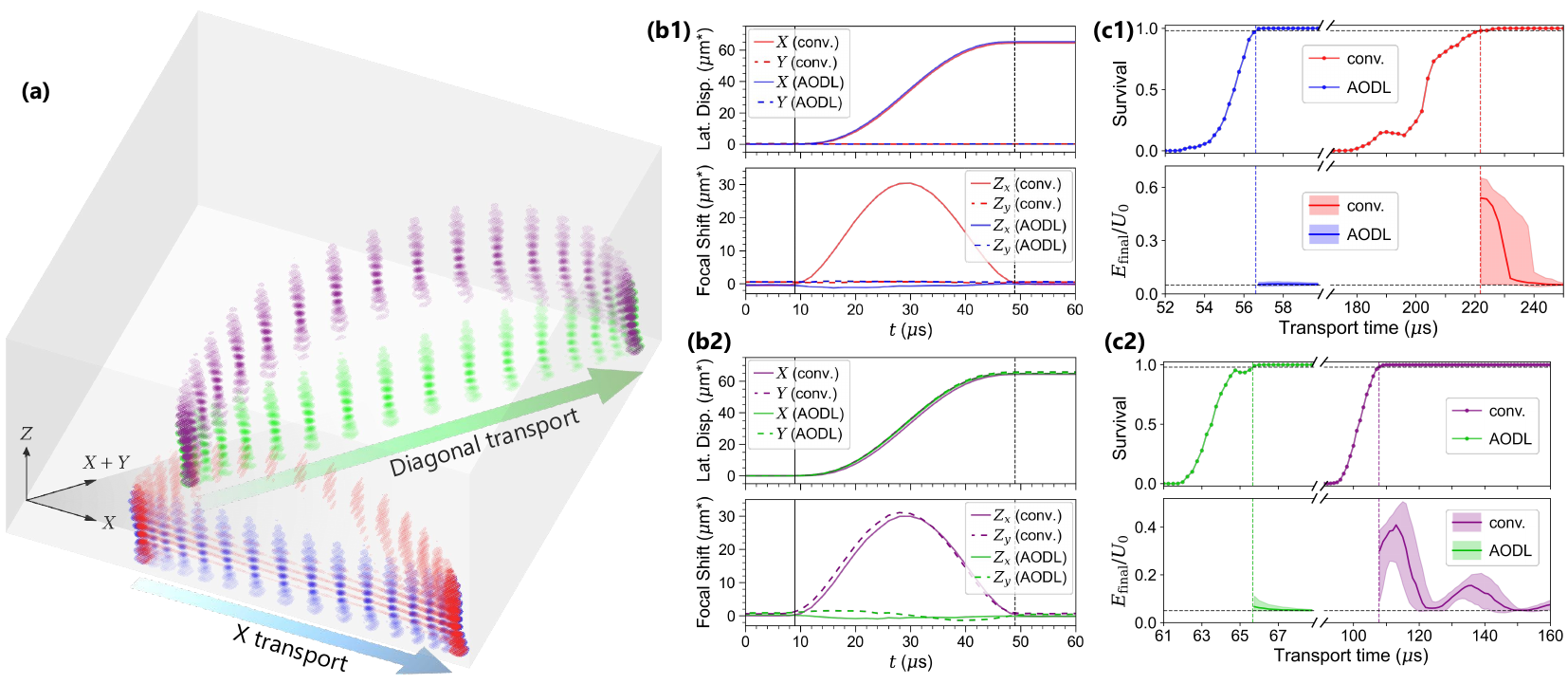}
    }
    \caption{
   \textbf{Comparison of conventional VS 3D-AODL tweezer trajectories.} 
   (a) \textit{3D reconstruction of minimum-jerk trajectories.} Each trajectory is a stroboscopic overlay of frames spaced 2$\,\mu$s apart, Using opacity to represent the intensity . Red/Purple: Conventional 2D-AOD transports along the +X and diagonal +(X+Y) directions, exhibiting out-of-plane focal shifts. Blue/Green: AODL transports along the same directions, which remain in-plane. 
   (b1, b2) \textit{Lateral displacement} (top) and \textit{focal shift} (bottom) versus time for +X (b1) and diagonal (b2) transport. In conventional cases, focal shifts are proportional to the time derivative of the corresponding lateral displacement.
    (c1, c2) \textit{Simulated atom survival probability and final temperature.} Horizontal dashed lines denote initial atom energy, shaded regions denote the 68\% confidence interval across Monte Carlo trajectories. Simulation parameters: $^{87}$Rb atoms, trap depth $U_0 = 2\pi\times 20$ MHz, initial energy $E_\text{init} = U_0/20$.
    \label{fig: M2}
    }
\end{figure*}

\paragraph{\textbf{Single AOD }}
To elucidate the origin of optical aberrations in traveling tweezers, first consider a single AOD driven by a linearly chirped tone, $f_{Ay}(t) = f_0+\beta\,t$ (Fig.~\ref{fig: M1}(b)) where $\beta$ is the  chirp rate. At each time slice, the acoustic wavevector $K(y)$ varies across the active aperture as $K(y) = K(0) + (2\pi\,\!\beta/v^2)  y$, with $v$ being the speed of sound in the acousto-optical crystal. The spatial gradient of the diffraction wavevector is then imprinted on the first-order deflection of the beam, creating an effective cylindrical lens with a dioptric power of $P_y = \lambda\beta/v^2$. Generalizing to nonlinear chirps, the cylindrical dioptric power and astigmatic interval are
\[ P_y = \frac{\lambda }{v^2}\dot{f}_{Ay}(t), \quad \Delta F =  \frac{\lambda F^2}{v^2}\dot{f}_{Ay}(t), \]
where $\lambda$ is the wavelength of light, and $F$ is the focal length of the final objective used to focus the deflected beam onto the image plane. To quantify astigmatism, we define the astigmatism factor $\sigma_\text{astig} = \Delta F/z_R$ where $z_R$ is the Rayleigh range. For example, when $\sigma_\text{astig} = 2$, the trap depth reduces by half and the maximum axial trapping force reduces by 62\%. 

\paragraph{\textbf{2D-AOD -- diagonal motion}}
Many neutral-atom platforms use two AODs crossed at 90$\degree$ (2D-AOD) to generate 2D tweezer motion in the lateral plane. A diagram of two 2D-AODs in a 4\textit{f} configuration is shown in Fig.~\ref{fig: M1}(a). The left inset depicts a linear chirp on AOD-$Ay$, which induces cylindrical lensing along the y-axis, as described above. Likewise, a linear chirp on AOD-$Ax$ induces cylindrical lensing on the x-axis.
The impact of cylindrical lensing on tweezer trajectories is shown in Fig.~\ref{fig: M2}(a).
When solely chirping AOD-Ax in a minimum-jerk ramp, the red trajectory appears to split into two branches -- the focus in x rising out of plane to a zenith about 30$\,\mu\text{m}^*$ above the static focal plane, and the focus in y remaining in-plane -- revealing severe astigmatism. In contrast, chirping both AOD-Ax and AOD-Ay with identical minimum-jerk ramps produces the purple trajectory, which reaches the same out-of-plane height but with 
no astigmatism.
The absence of astigmatism in diagonal motion is due to equal cylindrical lensing in the x- and y-axis,
which together act as a spherical lens.  Experiments have shown that such diagonal motion can be realized with higher transport velocities while maintaining trap stability and atom survival~\cite{Manetsch2025}.
Nevertheless, the out-of-plane motion -- along a direction in which the tweezer-trap curvature is weakest -- does place unwanted limits on the transport time.


\paragraph{\textbf{3D-AODL -- in-plane motion}} The assembly of four AODs into a 3D-AODL allows for cancellation of astigmatism and offers independent control over the three-dimensional position of the focal point. We use a 4f imaging configuration in which two sets of 2D-AODs (using +1 order deflection) are placed at the same orientation, as shown in Fig.~\ref{fig: M1}(a). The 4f relaying effectively inverts the first set of 2D-AODs (labeled $Ay$ and $Ax$) and overlays them onto the second set of 2D-AODs (labeled $By$ and $Bx$), forming counter-propagating pairs of sound waves: $Ax$ and $Bx$ ($Ay$ and $By$). 
If we counter-chirp each pair ($Ax$ and $Bx$, or $Ay$ and $By$) of AODs such that their frequency sum remains constant, the frequency difference maps to the beam steering angle or the lateral position of the tweezer (Fig.~\ref{fig: M1}(b)). At the same time, the lensing effects are strictly canceled so the tweezer focus remains in-plane and astigmatism-free (Fig.~\ref{fig: M2}: blue and green). The Monte Carlo simulation shown in Fig.\ \ref{fig: M2}(c1) and (c2) show that high-survival-probability translation can be realized in these lensing-free translated tweezers for shorter transport times as compared to the limits seen for systems in which the AOD lensing is uncompensated. In addition, motional heating caused by these in-plane translations drops much faster with increasing transport times, compared to conventional methods.

\paragraph{\textbf{3D-AODL -- omnidirectional motion}} Fig.~\ref{fig: M1}(c) shows that if we co-chirp each pair ($Ax$ and $Bx$, or $Ay$ and $By$) of AODs so that their frequency difference remains constant, the time derivative of the frequency sum maps to the corresponding axial focal position, while the lateral position does not change in time. 
Because the two pairs of AODs control the corresponding induced cylindrical lenses independently, maintaining $\Delta F = Z_x - Z_y = 0$ imposes a constraint on the chirp rates of the four AODs. 
This reduces the independent degrees of freedom of a 3D-AODL from four to three, matching the 3D motional degrees of freedom  (Tab.~\ref{tab: M1}).

\begin{table}[b]  
\centering
\begin{tabular}{ |c|c|c| } 
 \hline
 X-position & $ X =   \frac{\lambda F}{v}\big( f_{Bx} - f_{Ax} \big)$  \\ 
 \hline
 Y-position & $ Y =   \frac{\lambda F}{v}\big( f_{By} - f_{Ay} \big)$ \\ 
 \hline
 Z-position & $ \overline{Z} =  \frac{1}{2} \frac{\lambda F^2}{v^2}\big( \dot{f}_{Ax} + \dot{f}_{Bx} + \dot{f}_{Ay} + \dot{f}_{By} \big)$\\ 
 \hline
 Astigmatic interval & $\Delta F =  \frac{\lambda F^2}{v^2}\big( \dot{f}_{Ax} + \dot{f}_{Bx} - \dot{f}_{Ay} - \dot{f}_{By} \big)$ \\ 
 \hline
\end{tabular}
\caption{\label{tab: M1} Tweezer parameter vs AOD frequencies.  }
\end{table}

One can readily generalize the 3D motions above to  an array of astigmatism-free tweezers   by sending a superposition of multiple frequency tones onto AOD-$Bx$ and AOD-$By$, while keeping AOD-$Ax$ and AOD-$Ay$ single-tone. 


\section{Unconstrained 3D tweezer motion}

A significant constraint of the aforementioned methods is that any prolonged z-offset requires a continuous unidirectional chirp on all 4 AODs. Because AODs have a limited diffraction bandwidth from $f_\text{min}$ to $f_\text{max}$, determined by the Bragg condition~\cite{KORPEL197271}, the axial displacement of a tweezer formed by a single chirped-tone drive is constrained as 
\begin{equation}
    \left|\int_{t_0}^{t_1} Z(t) \text{d}t\right| < \frac{\lambda F^2}{v^2}(f_\text{max} - f_\text{min}).
\end{equation}
This constraint is illustrated in the bottom panel of Fig.~\ref{fig: M3}(a), where the diffraction efficiency rolls- off sharply near the edges of the band.  

One solution is to jump the frequencies of x-(y-)AOD pairs whenever one of them hits a frequency bound, so that the chirp may continue with a  \emph{new}  tweezer at the same spatial location~\cite{Kirkby2010, Rozsa2013}. However, such frequency jumps will inevitably lead to highly distorted acoustic waves running across the AOD aperture, distorting the optical wavefront and disturbing the tweezer-trapped atom, unless the light is blinked off until the acoustic wave front of frequency discontinuity has propagated through the active aperture. While the switching off of light is allowed in some occasions~\cite{hwang2023optical}, extended off-time of tweezers will lead to heating and loss of trapped atoms~\cite{Wang2020_AtomEnsemblesTweezer,Kuhr2005_DephasingStandingWave}.

\begin{figure}
    {\centering
    \includegraphics[width=1\linewidth]{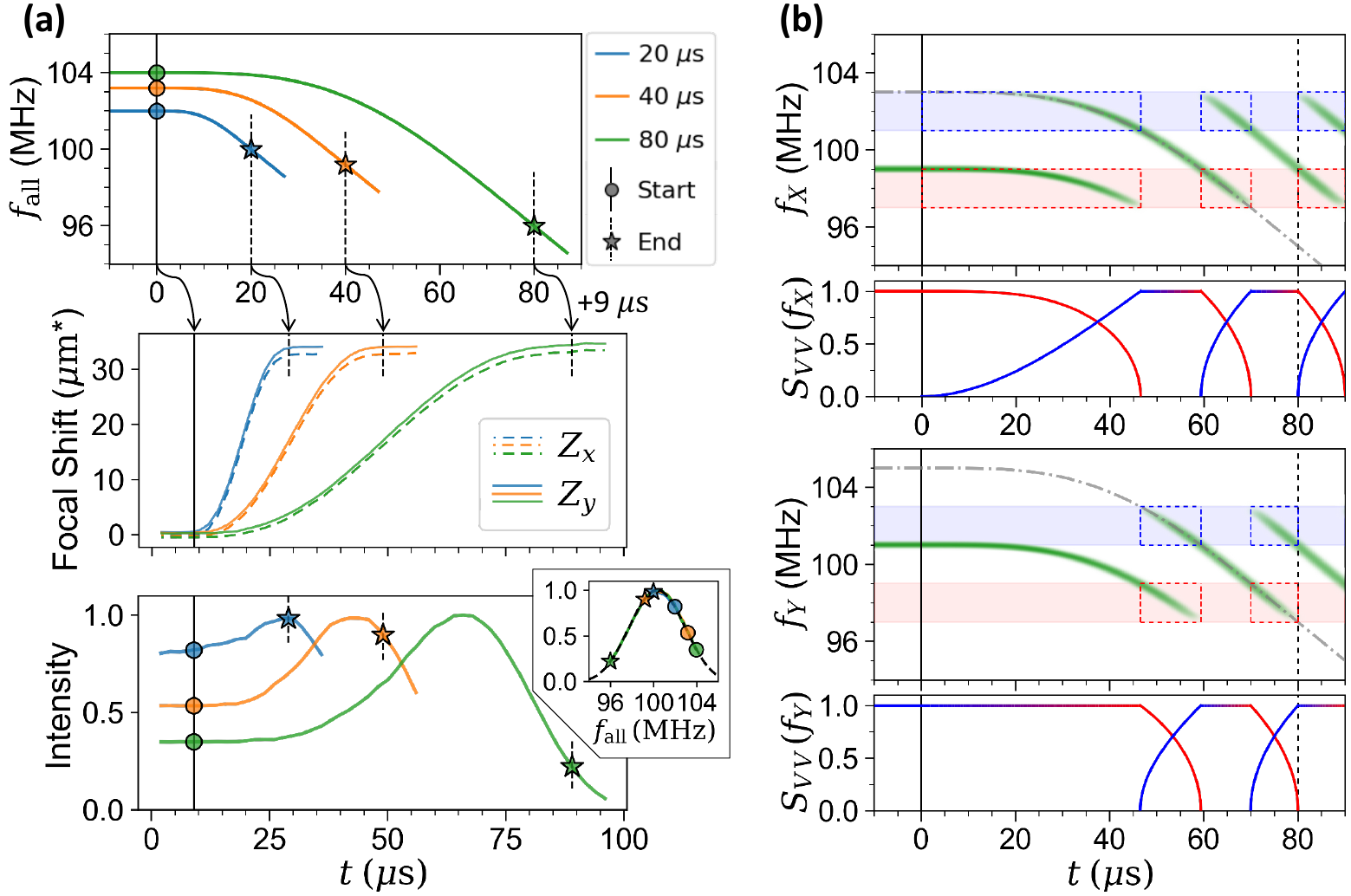}
    }
    \captionsetup{width=\linewidth} 
    \caption{
    \textbf{Constrained vs unconstrained axial transport.}
    (a) \textit{Single-tone minimum-jerk waveforms} with transport times of 20, 40, and 80 µs. Top: drive frequency vs time. Middle: focal shifts $Z_x$ and $Z_y$ remain nearly equal, indicating axial motion with minimal astigmatism. Bottom: tweezer intensity vs time, showing increasing losses for longer transports due to AOD diffraction efficiency roll-off, as shown in the inset.
    (b) \textit{fading-Shepard waveform} for an 80 µs transport. Top two: AODs $Ax $ and $Bx$. Bottom two: AODs $Ay$ and $By$. In each pair of plots, the top plot is a spectrogram of chirped rf tones (green) with single frequency extensions for reference (gray dash-dotted). The lower plot shows the power spectral density (PSD) of each tone. Blue (red) spectrogram regions indicate rf tones fading in (fading out), where  PSD  increases (decreases). 
    \label{fig: M3}
    }
\end{figure}
Our solution, inspired by the famous Shepard tones, is to fade in and fade out frequency tones to create a family of perpetual chirping waveforms -- the \emph{fading-Shepard} waveforms. As one tone fades out and the next tone fades in, a gradual transfer of the trapped atom from the \emph{old} to the \emph{new} tweezer occurs such that the total intensity remains unchanged. The optical interference between the \emph{old} and \emph{new} tweezers creates intensity modulation at the optical beatnote frequency, which can be in the MHz range and is, thus, much higher than the atom trap frequencies.  As such, trapped atoms will be unperturbed by such modulation. To avoid zero-frequency interference, we interlace the fading zone of the X-AOD pair and the Y-AOD pair so that they alternate, as shown in Fig.~\ref{fig: M3}(b). (Supplement, Sec. 1C.2).

A fading-Shepard $V_{\mu}(t)$ waveform can be parametrized using the number of tones $M$ (yielding $M+1$ tones during fading),  fading order $p$, and fading offset $\xi$.  In  Tab.~\ref{tab: M2}, we denoted as configuration $(M,  p, \xi )$. 
The fading order $p$ controls the power scaling of the fading amplitude in Eq.(\ref{eq: fading-power}), in order to maintain a constant total intensity throughout fading. The fading offset $\xi$ sets the relative timing between fading zones, allowing the x- and y-fading processes to be temporally interlaced. The waveform is given by
\vspace{-4pt}
\begin{align}
V_{\mu}(t) \!&=\!\!\! \sum_{n=-\infty}^{\infty}\! \!\! A_{\mu}^{(n)}(t)
   \cos\!\left[ 2\pi\!  \int_0^t 
   \!\big(f_{0} \!+\! f_{\mu,\text{lat}}(\tau) \!+\! f^{(n)}_{\mu, Z}(\tau)\big)\, \mathrm{d}\tau \right. \nonumber \\
 &\left. \quad +\, \phi_{\mu}^{(n)} \right],
\; \mu \in \{Ax, Ay, Bx, By\}.
\end{align}

Lateral motion is generated by $f_{\mu,\text{lat}}(t) = \mp \frac{v}{2\lambda F} X(t)$ for $\mu = Ax, Bx$ and  $f_{\mu,\text{lat}}(t) = \mp \frac{v}{2\lambda F} Y(t)$ for $\mu = Ay, By$. 
Axial motion is generated by $ f^{(n)}_{\mu, Z}(t) = \frac{v^2}{2\lambda F^2} \int_0^{t} Z(\tau)\text{d}\tau  + \left(n+ \xi_{\mu} \right) \Delta f  $, where the frequency spacing $\Delta f$ controls tweezer spacing. To interlace x- and y-fading zones we set $\xi_{Ax} = \xi_{Bx} = 0$ and $\xi_{Ay} = \xi_{By} = 0.5$. The initial frequencies and phases are set by $f_0$ and $\phi_{\mu}^{(n)}$.

\begin{figure*}
    \includegraphics[width=1\linewidth]{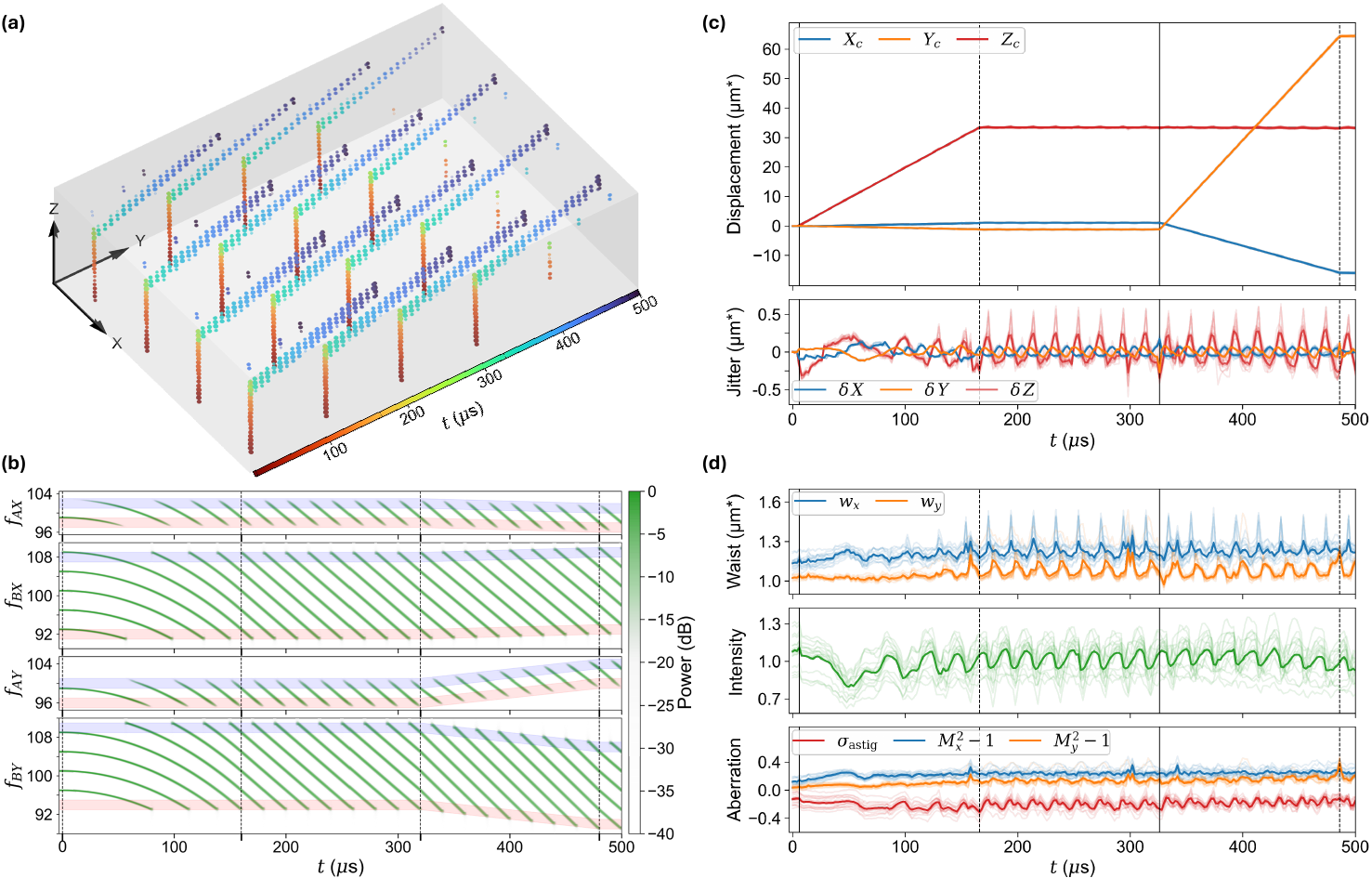}
    \caption{
    \textbf{Out-of-plane ``L'' shaped trajectory of a 4$\times$4 tweezer array. }   
    (a) \textit{Reconstructed 3D tweezer trajectories,} color-coded by time, showing an “L”-shaped path of a uniformly spaced array.
    (b) \textit{Spectrogram of fading-Shepard waveforms}. The waveform is divided into three segments: linear motion in Z (0–160 $\mu$s), static with constant Z offset (160–320 $\mu$s), and linear motion in X with fixed Z offset (320–480 $\mu$s).
    (c) \textit{Position stability.} Top: X, Y, and Z trajectories; Bottom: Deviation from the ideal path, showing  axial jitter$<$0.4\,$\mu\text{m}^*$ and lateral jitter$<$0.08\,$\mu\text{m}^*$. 
    Array spacing of $4$ MHz corresponds to $32.5\,\mu\text{m}^*$. 
    (d) \textit{Shape stability:} waists, intensity, and aberration metrics remain stable, with intensity fluctuating by $\sim\pm$9\% and astigmatism by $\sim\pm$0.1.
    \label{fig: M4}
    }   
\end{figure*}

The amplitude of each frequency tone $A^{(n)}_{\mu}$ is faded in and out in time -- for example, the upper (blue) and lower (red) fading zones are shown in Fig.~\ref{fig: M3}(b). 
Each frequency tone takes full (zero) amplitude between (outside) fading zones, and smoothly fades in and out within fading zones:  
\begin{equation}
\label{eq: fading-power}
A^{(n)}_{\mu}= 
\begin{cases}
1,\quad \quad    \quad \quad  \quad     \quad \left|f_{\mu, Z}^{(n)}\right| \le \dfrac{(M-\eta)\,\Delta f}{2}, &\\[5pt]
0, \quad  \quad   \quad    \quad  \quad  \quad \left|f_{\mu, Z}^{(n)}\right| \ge \dfrac{(M+\eta)\,\Delta f}{2}, & \\[5pt]
\cos^{\,p_\mu}\!\left[ \frac{\pi}{2\eta} \Big( \frac{| f_{\mu, Z}^{(n)}|}{\Delta f} - \frac{M}{2} \Big) + \frac{\pi}{4} \right], \;\text{otherwise,}&
\end{cases}
\end{equation}
where $0<\eta\le1/2$ is the fading duty, defined as the spectral width of the fading zone divided by the frequency spacing, and is set to $\eta=1/2$ 
in this work. 
The fading orders need to satisfy $p_{Ax}+p_{Bx}=p_{Ay}+p_{By}=1$, 
    ensuring that the sum of \emph{old} and \emph{new} tweezer powers remains constant.

\begin{table}[b]
\begin{center}
    \begin{tabular}{|c|c|c|}
    \hline
    $(M,\;  p,\; \xi\, )$  & Single tweezer  & $M_x\times M_y$ array \\
    \hline
    \hline
    AOD-$Ax$& $(1,\; 0.5,\;  \;0\;\;)$ & $(\;1\;\;,\;  1,\; \;0\;\;)$ \\
    \hline
    AOD-$Ay$& $(1,\;  0.5,\; 0.5)$ & $(\;1\;\;,\; 1,\;  0.5)$ \\
    \hline
    AOD-$Bx$& $(1,\; 0.5,\;\;0\;\;)$ & $(M_x,\; 0,\; \;0\;\;)$ \\
    \hline
    AOD-$By$ & $(1,\; 0.5,\; 0.5)$ & $(M_y,\; 0,\; 0.5)$ \\
    \hline
    \end{tabular}
\end{center}
    \caption{Configuration of fading-Shepard waveform with interlaced fading. 
    }
    \label{tab: M2}
\end{table}

An important consideration for fading–Shepard waveforms is the relative phase of each tone, especially when the AOD operates near saturation, where third-order intermodulation (IM3) between tones is significant~\cite{Endres2016, Gazalet1993AO}. To suppress IM3, we adopt a generalized Schroeder-phase scheme \cite{Schroeder1970}, where the initial phase for the $n$-th tweezer in the x(y) direction is:
\begin{equation}
    \phi_{Bx(y)}^{(n)}=2 \pi \times\frac{n(n-1)}{2M_{x(y)}}  
\end{equation}

Compared to a random phase offset, the choice of the Schroeder phase minimizes the crest factor of the overall waveform \cite{Shibasaki2020}. We anticipate the Schroeder phase to provide similar performance to other IM3-suppressing phases such as the Kitayoshi, Narahashi, or Newman phases. In addition, tweezer intensity uniformity across the generated array may be attained with iterative camera feedback methods \cite{MittenbuhlerAOD2025,TrypogeorgosAOD2013}.

\section{Benchmarking tweezer trajectories}

To characterize the stability, span and speed of the tweezer array, we implemented a fading-Shepard waveform on the 3D-AODL to steer a $4\times4$ tweezer array along a designed ``L''-shaped trajectory. As shown in Fig.~\ref{fig: M4}, the sequence consists of three consecutive $160~\mu$s segments: (i) a quadratic chirp on all four AODs that produces a $34~\mu\text{m}^*$ linear displacement along $Z$; (ii) a linear chirp on all four AODs that holds the tweezers at a constant $Z$ offset; (iii) fading-Shepard chirps on all four AODs with an opposite-signed slope added to each AOD pair, yielding a lateral translation of $65~\mu\text{m}^*$ in $Y$ and $-16.25~\mu\text{m}^*$ in $X$ while maintaining the constant $Z$ offset. Reconstructed trajectories confirm that all 16 traps follow the programmed ``L''-shaped path with high uniformity across the array. Trap-shape analysis indicates that waists, intensity, and aberration remain stable throughout the motion, with intensity fluctuations of $\pm 9\%$, astigmatism factor variations of $\pm 0.1$ (comparable to the average astigmatism factor of $-0.1$ set by optical alignment imperfections), and an average $M^2$ increase of 10\%. Displacement analysis further shows that the tweezers track the ideal path with high fidelity, exhibiting axial jitter $<0.4~\mu\text{m}^*$ and lateral jitter $<0.08~\mu\text{m}^*$, as shown in the bottom part of Fig.~\ref{fig: M4}(c).

We attribute the position jitter and beam waist (and thus intensity and $M^2$) fluctuations to three independent sources and provide corresponding solutions. \textit{Firstly}, if there is an unequal distance between the piezoelectric transducer and the beam on a counter-propagating (x- or y-) AOD pair, the fading-Shepard waveform will exhibit a temporal offset leading to both intensity fluctuation and position jitter of the tweezer during fading. This effect is compensated for by manually adding a suitable delay between the rf waveforms that are sent to the two same-axis AODs (Supplement, Sec. 3B). 
\textit{Secondly}, 
axial offsets in the $4f$ imaging system between the AODs and the two conjugate planes can cause the magnification to be $M\neq-1$. This magnification mismatch effectively induces velocity mismatch on a counter-propagating pair, which we compensate for by linearly dilating the frequency range of AOD-Bx and -By with a small factor. In principle, this effect could also be compensated for with a $4f$ imaging system of larger $f$ or slimmer AODs. \textit{Lastly}, the limited first AOD bandwidth and the acoustic-irising effect~(Supplement, Sec.~3) can induce intensity fluctuation that's synchronized to the fading cycle. This effect can be compensated for by introducing rf power compensation factors during fading. 

As shown in Fig.~\ref{fig: M5}(a, b), the effective range of the tweezer motion spans at least 130$\,\mu\text{m}^*$ ($\sim$125 times the beam waist) in the lateral dimensions, and 136 $\mu\text{m}^*$ ($\sim$32 times the Rayleigh range) in the axial dimension. The size of the tweezer array itself may span a larger range ($> 200\,\mu\text{m}^*$) in the lateral dimensions (Fig.~\ref{fig: M5}(d)), and is limited in our demonstrations by the camera sensor size. We did not benchmark the absolute maximum velocity of the tweezers during long-range transport, as no visible aberration was observed in the in-plane transport paths (Fig.~\ref{fig: M2}) where peak velocity during transport reached $4.2~\mathrm{m/s}$; this is already faster than the typical maximum transport speed used for long-range transport ($\sim$ 0.5 m/s). Though we predict that higher-order aberrations exist in the form of $n^\mathrm{th}$ order Zernike polynomials proportional to the $(n\!-\!1)^\mathrm{th}$ order time derivatives of the AOD frequency (for example, the comatic aberration scales as $\frac{\mathrm{d}^2 f(t)}{\mathrm{d}t^2}$, which is proportional to the lateral acceleration), these higher-order effects are unlikely to limit transport times in a typical neutral atom experiment, where trap acceleration and higher-order time derivatives are small.

Finally, we note that nearly all parameters described above scale favorably with the AOD active aperture $D_\mathrm{AOD}$. For a fixed numerical aperture (NA), a larger aperture allows either a longer-focal-length objective or a relay telescope that images the larger AOD output onto the smaller objective input, effectively reducing the acoustic velocity on the image plane. In both cases, $F/v \propto D_\mathrm{AOD}$, so the lateral (axial) displacement scales linearly (quadratically) with $D_\mathrm{AOD}$ (see Tab.~\ref{tab: M1}). Conversely, to achieve the same axial shift $|\Delta Z|$, the required chirp rate scales as $|\beta| \propto D_\mathrm{AOD}^{-2}$. A slower chirp rate relaxes alignment tolerance and lengthens the fading cycle, thereby mitigating acoustic-irising effects.

\begin{figure*}
    {\centering
    \includegraphics[width=1\linewidth]{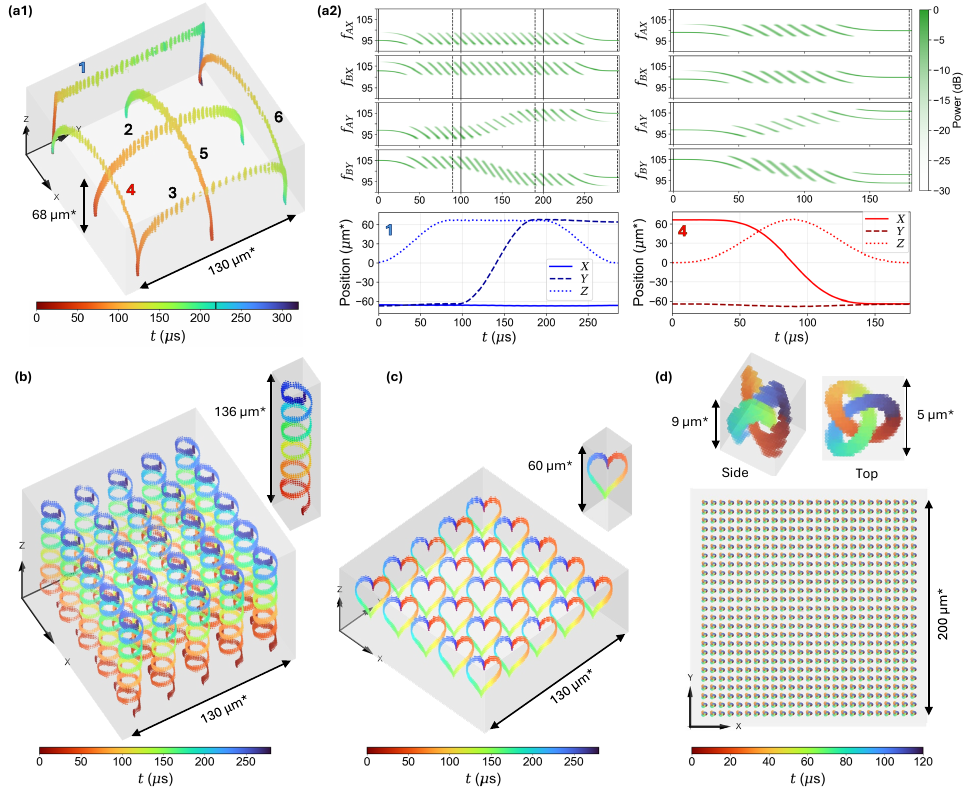}
    }
    \caption{\textbf{Programmable 3D trajectories.} 
    (a1) \textit{Elevated lateral transport.} Each trajectory is designed to lift an atom out-of-plane by 68\,$\mu\text{m}^*$, followed by a 130\,$\mu\text{m}^*$ lateral translation before returning to the original plane. 
    (a2) Drive spectrograms (top) and tweezer positions (bottom) for representative cases: trajectory 1 is constructed from three stitched minimum-jerk segments, and trajectory 4 is a single smooth path that minimizes overall jerk. 
    (b--c) \textit{Programmable trajectory shaping} in a 5$\times$5 array with 32.5$\,\mu$m$^*$ spacing. Shown here are helical and heart-shaped paths, illustrating flexible waveform programmability. 
    (d) \textit{3D potential modulation} in a 25$\times$25 array  with 8.125$\,\mu$m$^*$ spacing. Omnidirectional motion enables complex periodic modulation of trap potentials per site, such as the \emph{Trefoil-knots} as shown. No fading-Shepard waveforms are required in this case due to the small modulation amplitude.}
    \label{fig: M5}
\end{figure*}

\paragraph{Elevated Lateral Transport} Translating atoms or other particles along 3D trajectories is desired for various applications~\cite{Grier2003, Berthelot2014}, including the assembly of a defect-free 3D atom array~\cite{Barredo2018,  Lee2016}, and rearranging atoms trapped in a 2D optical lattice. 
For example, recent experiments have made use of single atoms trapped at the interference of cylindrically focused optical beams \cite{Schlosser2023, Gyger2024}, in which 
AOD-generated optical tweezers translate atoms within optical lattices. 
If the translation is performed purely in the transverse plane, the translated atom will be influenced by the washboard potential through which it is translated, generating a rapidly varying force on the atom, inducing heating\cite{Weitenberg2011} and potentially loss.

To avoid lattice-induced heating, we propose a class of optical tweezer transport trajectories that elevate the atom above the lattice plane, thereby mitigating the constraints of periodic confinement during rearrangement and providing access to multi-layer structures. In Fig.~\ref{fig: M5}(a1), we demonstrate six such elevated trajectories covering a square area. Trajectory 1 demonstrates uncoupled 3D motion, while trajectories 2 through 6 demonstrate smooth control of out-of-plane movements. We analyze trajectories 1 and 4. Trajectory 1 consists of three segments: (i) a fading-Shepard chirp lasting $90~\mu$s on all four AODs that produces a $68~\mu\text{m}^*$ minimum-jerk ascent along $Z$, followed by a static hold of $10~\mu$s. (ii) In addition to fading-Shepard chirps, frequency ramps are added to AODs $Ay$ and $By$, yielding a  $90~\mu$s lateral minimum-jerk translation of $130~\mu\text{m}^*$ in $Y$, followed by a static hold of $10~\mu$s. (iii) a fading-Shepard chirp lasting $90~\mu$s on all AODs that produces a $-68~\mu\text{m}^*$ minimum-jerk descent along $-Z$ (Fig. \ref{fig: M5}(a2, left)). Trajectory 4 is a single smooth motion: fading-Shepard chirps ramping up and then down in $Z$ lasting $180~\mu$s on all AODs, with frequency ramps added to AODs $Ax$ and $Bx$ to produce a $130~\mu\text{m}^*$ final displacement along $X$. The reconstructed trajectories of Trajectory 1 and Trajectory 4 confirm that tweezer motion is stable and follows the programmed paths. The large axial displacements of trajectories ($68\,\mu\text{m}^*$, 16 times the Rayleigh range) allow for overhead transport of selected atoms without disturbing atoms below.

\section{Discussion and Outlook}
The results presented show that speed limits in AOD-based atom transport are set primarily by chirp-induced acoustic lensing, rather than by intrinsic atomic or optical constraints. Using fading-Shepard waveforms, the 3D-AODL enables astigmatism-free, omnidirectional tweezer array motion at velocities beyond those accessible with conventional chirped-AOD approaches, enabling out-of-plane rearrangement paths. During the completion of this work, alternative solutions to the issue of acoustic lensing emerged \cite{picard2025, guo2025AOD3D}. The 3D acousto-optic deflector exhibited in \cite{picard2025} provides a highly compact device that enables sustained focal offset without the need for a fading-Shepard waveform, but it is designed for spherical lensing, so it would need to be used in series with the 3D-AODL to cancel astigmatism. An alternative device exhibited in \cite{guo2025AOD3D} provides rapid access to 3D using retroreflection, with the additional advantage of a wider rf bandwidth by leveraging the anisotropic Bragg condition on both passes, and the additional optical power loss due to double-passing. The single tweezer fading-Shepard waveforms can be readily utilized on this device to enable sustained focal offset.  

Our 3D-AODL presents additional control parameters due to the need for four independently programmed waveforms and the introduction of fading-Shepard waveforms.  For in-plane rearrangement, astigmatism-free motion is achievable with no added computational cost relative to conventional 2D AODs. For true-3D applications of the 3D-AODL, we anticipate the bulk of the computational time will still be image readout and CPU processing of the optimal motion sequence, similar to the current paradigm for 2D AODs \cite{Guo2025movingtweezerfpga, wang2023fpga}. 

Furthermore, the 3D-AODL enables a variety of new applications. As a standalone device, the 3D-AODL functions both as an astigmatism-free beam deflector and a varifocal lens. By accessing the third dimension, it enables rapid in-plane and \emph{inter-plane sorting} in three-dimensional atom arrays~\cite{Barredo2018,Schlosser2023} as well as facilitating dynamic rearrangement within a 2D lattice~\cite{Norcia2024, Tao2024, Gyger2024} via \emph{elevated lateral transport} (Fig.~\ref{fig: M5}(a)). Other applications include: \emph{Atom-chain assembly}  (Fig.~\ref{fig: M5}(b,c)) via 1D parametric motion along prescribed 3D paths (e.g., helices), enabling rapid, defect-free construction of ordered chains and more intricate geometries~\cite{Peter2024}. \emph{3D potential modulation} (Fig.~\ref{fig: M5}(d)) along a periodic trajectory, providing a practical route to Floquet engineering of time-dependent lattice Hamiltonians~\cite{Eckardt2017} and tweezer arrays~\cite{ Grochowski2025}.  In conjunction with a static spatial light modulator, our system may be used to modulate large tweezer arrays with single-site addressability~\cite{Zhang2024, Radnaev2025}. Beyond atomic physics applications, the improved spatiotemporal wavefront control made possible with our fading-Shepard waveforms enables broader utility in microscopy~\cite{Soto:18, Kirkby2010}, imaging~\cite{Katona2012}, and scanning~\cite{Rozsa2016}

. 

\section{methods}

\paragraph{\textbf{Unit Conversion}}
Spatial coordinates are reported in converted units, $\mu\text{m}^*$, to reflect typical atom trapping scales. Tweezer tomography measured with a $F=100$~mm imaging lens is rescaled to an effective $F^*=6.5$~mm objective (as in the Monte Carlo simulation). This corresponds to multiplying the lateral positions (pixel index $\times$ $3.45\,\mu$m) by $0.065$ and the axial positions (translation stage) by $0.065^2 = 4.23\times10^{-3}$. Using converted units, the static tweezer has a waist radius of 1.1(1) $\mu\text{m}^*$ and a Rayleigh range of 4.2(3) $\mu\text{m}^*$, and $1\,$MHz of frequency difference maps to 8.125 $\mu\text{m}^*$ spacing.

\paragraph{\textbf{Monte Carlo Simulation}}
Simulation parameters: $^{87}$Rb atoms, trap depth $U_0 = 2\pi\times 20$ MHz, initial energy $E_\text{init} = 2\pi\times 1$ MHz with randomized positions and velocity directions; wavelength $808$ nm; objective $F^* = 6.5$ mm, NA $=0.5$, effective NA $=0.3$ (set by input beam radius).

\paragraph{\textbf{Experimental apparatus}}
A 808-nm diode laser (Thorlabs LD808-SEV500) is shuttered by an AOM (IntraAction ATM-2701A2) for stroboscopic imaging. The shuttered output is fiber-coupled and sent through the 3D-AODL, which consists of two 2D-AODs (AA Optoelectronics DTSXY-400-800.860 and DTSXY-400-780-002), $4f$-relayed by two $150\,$mm Hastings triplets (Fig.~\ref{fig: M1}(a)) and driven by two synchronized dual-channel AWGs (Spectrum M4i.6631-x8). Imaging is performed with a $100\,$mm doublet on a CMOS camera (Thorlabs Zelux) mounted on a motorized translation stage (Thorlabs MTS50-Z8). 


\paragraph{\textbf{Data Acquisition}}
Each experimental run produced a 4D dataset ($X$, $Y$, $Z$, $t$), with each element storing a single pixel intensity. Images were taken at delayed time steps (250-ns shutter, SRS DG535) as the translation stage was stepped from $Z_\text{min}=-24.5$ mm to $Z_\text{max}=24.5$ mm.
The camera exposure is synchronized to the waveform period so each image integrates over a fixed number of shuttered pulses.

\section{Acknowledgments}

We thank Nathaniel B Vilas, Hannah J Manetsch, Elie Bataille, Xudong Lv, and Mark J Stone for fruitful discussions. We acknowledge support from the AFOSR (Grant No.\ FA9550-1910328), from ARO through the MURI program (Grant No.\ W911NF-20-1-0136), from DARPA (Grant No.\ W911NF2010090), from the NSF (QLCI program through grant number OMA-2016245), and from the U.S. Department of Energy, Office of Science, National Quantum Information Science Research Centers, Quantum Systems Accelerator. J.H. acknowledges support from the Department of Defense through the National Defense Science and Engineering Graduate (NDSEG) Fellowship Program.


\textbf{Competing interests:} Y.H.L., N.S. and D.M.S.K. are inventors on a patent application related to the 3D-AODL and fading-Shepard technology reported in this work. The authors declare no other competing interests.

\textbf{Data and materials availability:} Experimental data and simulation code are available from the corresponding author upon reasonable request.

\endgroup
\clearpage
\clearpage
\onecolumngrid


\captionsetup[figure]{justification=capjust, singlelinecheck=true, width=\textwidth}
\captionsetup[table]{justification=capjust, singlelinecheck=true, width=\columnwidth}

\makeatletter
\AtBeginDocument{\let\LS@rot\@undefined}
\makeatother

\newcommand{\Tr}{\text{Tr}}

\newcommand{\pa}[1]{\partial_{#1}}

\newcommand{\smic}{\sigma^-}
\newcommand{\splc}{\sigma^+}
\newcommand{\szc}{\sigma^z}
\newcommand{\cs}{c^*}
\newcommand{\nn}{\bold{n}}
\newcommand{\rr}{\bold{r}}
\newcommand{\rrop}{\hat{\rr}}
\newcommand{\pp}{\bold{p}}
\newcommand{\ppop}{\hat{\pp}}
\newcommand{\xop}{\hat{x}}
\newcommand{\sz}{\hat \sigma^{z}}
\newcommand{\sx}{\hat \sigma^{x}}
\newcommand{\sy}{\hat \sigma^{y}}
\newcommand{\spl}{\hat \sigma^{+}}
\newcommand{\smi}{\hat \sigma^{-}}
\newcommand{\id}{\mathbb{1}}

\SetKwInput{KwData}{Input Devices}

\captionsetup[figure]{name=Fig., font=small}
\captionsetup[table]{font=small}
\frenchspacing

\pagenumbering{roman} 
\setcounter{page}{1}   

\begin{center}
{\bfseries\large Supplementary Material for:\\[2pt]
Astigmatism-free 3D Optical Tweezer Control for Rapid Atom Rearrangement}
\end{center}
\vspace{0.5em}

\tableofcontents
\markboth{}{}

\setcounter{page}{1}
\renewcommand{\thepage}{\roman{page}}

\setcounter{equation}{0}
\renewcommand{\theequation}{S\arabic{equation}}

\setcounter{figure}{0}
\renewcommand{\thefigure}{S\arabic{figure}}

\setcounter{table}{0}
\renewcommand{\thetable}{S\arabic{table}}

\section{Theoretical Model}
A full simulation of electromagnetic field evolution in acousto-optic material in the presence of an arbitrary acoustic field is extremely cumbersome and computationally demanding~\cite{KORPEL197271, chatterjee1990}. Therefore, we adopt two approximations to simplify the calculation while capturing the essential physics of light diffraction through multiple AODs. 

    In the theoretical analysis, it is assumed that we are operating in the weak-drive limit, i.e., the maximum optical phase modulation is much less than one. In the weak-drive limit, the optical field may be linearized in phase modulation, leading to a linear superposition of phase contributions and the suppression of higher-order sidebands and higher-order intermodulation terms. In reality, for considerations of laser power efficiency, AODs are generally driven beyond the weak drive limit, and special phase designs are of great importance to suppress third-order intermodulation (IM3) (Sec.~\ref {Schroeder}). 
    
\begin{figure}
    \centering
    \includegraphics[width=1\linewidth]{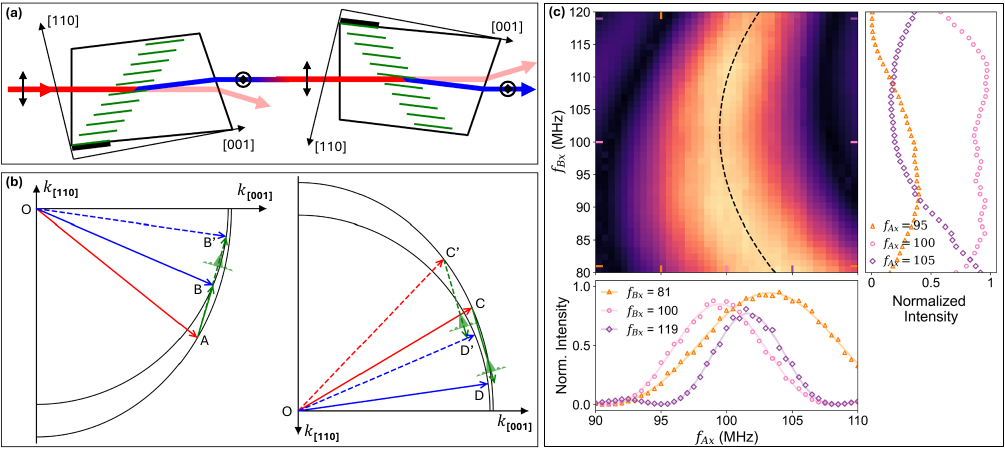}
    \caption{\textbf{Frequency dependent diffraction efficiency.}
    (a) \textit{Real-space diagram of a counter-propagating pair of AODs.} The principal axes and the surface cuts of the crystals are as shown. With respect to each AOD crystal, the incident and the non-deflected beams (red) have in-plane polarization (extraordinary ray), while the deflected beams (blue) have out-of-plane polarization (ordinary ray). The green parallel lines illustrate the wavefronts of the slow-shear mode acoustic wave with a walk-off from the transducers (bold black line segments).  (b) \textit{Momentum-space diagrams.} The left (right) panel plots the momentum space of the first (second) AOD. Within each panel, the outer (inner) curve shows the slowness surface of the extraordinary (ordinary) light. In the left panel, $\protect\vv{OA}$ is the k-vector of the incident light, and $\protect\vv{OB}$ ($\protect\vv{OB'}$) is the k-vector of the deflected light that corresponds to a lower (higher) acoustic frequency sound wave with the k-vector $\protect\vv{AB}$ ($\protect\vv{AB'}$). In the right panel, $\protect\vv{OC}$ ($\protect\vv{OC'}$) is the k-vector of the incident light that is simply the output k-vector of the first AOD driven at lower (higher) frequency $f_\text{Ax}$, and $\protect\vv{OD}$ ($\protect\vv{OD'}$) is the k-vector of the deflected light when the acoustic frequency of the second AOD $f_\text{Bx}$ is fixed. Perpendicular to each sound k-vector (green), a sinc function is drawn as a ruler for deviation from the Bragg phase matching condition; evidently, the phase matching condition is better satisfied in the first AOD than in the second AOD.   (c) \textit{ Measured overall diffraction efficiency.} The intensity of a single static tweezer is recorded while holding the $y$-axis drive frequencies constant ($f_\text{Ay}$ = $f_\text{By} = 100$ MHz) and scanning the $x$-axis frequencies, $f_{\text{Ax}}$ and $f_{\text{Bx}}$,
    across a 2D plane. The heatmap is shown with a power-law scaling (exponent of $1/2$) to enhance the visibility of the sinc-function fringes. The black-dashed line is a guide to the eye along the maximum-efficiency ridge; the bottom and right-hand panels provide 1D cross-sections of the 2D data.}
    \label{fig:anisotropic}
\end{figure}

    It is also assumed that there is no diffraction efficiency dependency on the drive frequency or the incident angle of the AODs. While this assumption is valid in the Raman-Nath regime, it does not hold in our case as we are deep in the Bragg regime with a narrow input angular aperture for the Bragg condition to be satisfied. In particular, the AODs that we use (AAOptic DTSXY-400-800) leverage the anisotropic birefringence of the $\text{TeO}_2$ crystal to expand the Bragg condition using the (near) tangential phase matching (TPM) condition \cite{Kim2008, Jiang2012, aaopto2013theory} for the acoustic k-vector. Experimentally, the 3D-AODL enjoys the TPM-broadened bandwidth on the drive frequency of the second AOD but not that of the first. This is because the acoustic frequency of the first AOD alters the incident angle onto the second AOD, as shown in Fig.~\ref{fig:anisotropic}.  However, as long as we operate the AODs near the flat regions near the center ridge we can assume a constant diffraction efficiency for first-order theoretical analysis. 

\subsection{Single-tone waveforms}
\subsubsection{Wavefront and aberration analysis}

\paragraph*{Single AOD} Suppose we have a Gaussian input beam, and we are using the first-order diffraction of the AOD. This AOD has acoustic waves traveling in the $+x$ direction, which generate a traveling wave of optical phase delay $\Psi(x,t) = \psi(t-x/v) = C\, V(t-x/v)$, where $v$ is the sound velocity, C is the phase-voltage ratio constant, and $V(t)$ is the drive waveform. The optical wavefront after the AOD can be expressed as a result of applying the pupil function on the incident waveform $U_\text{out}(x,y, t) = P(x,y, t) U_\text{in}(x,y, t)$:
\begin{equation} 
    P(x,y, t) = \exp(i\, \Psi(x,t)) = \exp(i\, C \, V(t-\frac{x}{v})) \simeq  \big( 1 + i\, C \, V(t-\frac{x}{v})\big)
\end{equation}
Where  the expansion of the wavefront modulation only goes up to $\pm 1$ orders. It is convenient to write the drive waveform in the rotating frame around the center resonant frequency $f_c$, so that $V(t) = A(t) \cos(2\pi f_c \,t + \phi(t))$, where $\phi(t)$ is the phase modulation of the drive waveform (not to be confused with $\psi(t)$ which is the optical phase delay). We now have:
\begin{equation}
    P(x,y, t) \simeq  1 + i\, C \,A(t-\frac{x}{v})   \cos\left((2\pi f_c \,(t-\frac{x}{v}) + \phi(t-\frac{x}{v})\right) = 1 + P^{(1)}(x,y, t) + P^{(-1)}(x,y, t)
\end{equation}
If we only look at the first-order diffraction,

\begin{align}
        P^{(1)}(x,y, t)\simeq \frac{iC}{2}   A(t-\frac{x}{v})  \exp\left(-i\,(t-\frac{x}{v}) ( 2\pi f_c  + \phi)\right)
\end{align}
The term $i(2\pi f_c \,\frac{x}{v})$ can be omitted because it signifies a constant deflection angle which can be absorbed into the redefinition of optical axis. The term $\exp\left(-2\pi i f_c \,t\right)$ can be also omitted because it signifies a constant optical frequency shift. Then the pupil function for the first-order diffraction wavefront can be Taylor expanded as:

\begin{align}
P(x,y,t) &\propto A \left(t-\frac{x}{v}\right) \exp\left(-i\phi\left(t-\frac{x}{v}\right)\right) \\
&= \Big(\underbrace{A(t)}_{\text{amp}} 
- \underbrace{\frac{A^{(1)}(t)}{v} x}_{\text{displacement}} 
+ \underbrace{\frac{A^{(2)}(t)}{2! v^2} x^2}_{\substack{\text{irising/}\\ \text{soft aperture}}} - \cdots\Big) \cdot\; \nonumber \\ 
& \qquad \exp\Big(\underbrace{-i\phi(t)}_{\text{phase}} 
+ \underbrace{i \frac{\phi^{(1)}(t)}{v} x}_{\text{deflection}} 
- \underbrace{i \frac{\phi^{(2)}(t)}{2! v^2} x^2}_{\substack{\text{lensing/} \\ \text{astigmatism}}} 
+ \underbrace{i \frac{\phi^{(3)}(t)}{3! v^3} x^3}_{\text{coma}} - \cdots \Big) \label{s5} \\
&\simeq \Tilde{A}(t)  \cdot
\exp\Bigg( \underbrace{2\pi i\, \frac{ f(t)}{v} x}_{\text{deflection}} 
- \underbrace{2\pi i\, \frac{ \dot{f}(t)}{2 v^2} x^2}_{\substack{\text{lensing/} \\ \text{astigmatism}}} \Bigg), \quad\quad \text{where} \;  \Tilde{A}(t) = A(t) e^{-i\phi(t)}
\end{align}

The terms on the left(right) modulate the amplitude(phase) of the wavefront. The zeroth-order amplitude modulation is a global term that controls the time-dependent power of the optical tweezer; the first-order amplitude modulation generates beam displacement before the objective, which is equivalent to tilt of tweezer angle after the objective without moving the tweezer position; the second-order amplitude modulation generates a soft aperture on the beam, sizing it down before the objective and widens the focal waist after the objective (Sec.~\ref{acoustic-irising}). The zeroth-order phase modulation is a global term that controls the optical phase/frequency of the optical tweezer; the first-order phase modulation generates beam deflection before the objective, which is equivalent to tweezer lateral displacement after the objective; the second-order phase modulation is effectively a cylindrical lens, which can either cause astigmatism of the tweezer or used to generate axial displacement after the objective; the third and higher order phase modulation generates higher-order aberrations such as coma. For simplicity, we will limit most of the discussion to only the zeroth-order amplitude modulation and up to second-order phase modulation. 

As an example, if we are driving the AOD with a fixed amplitude $A$ and a frequency that's linearly chirping in time, $f(t) = f_0 + \beta t$ (as the detuning from $f_c$), then the output wavefront $  U_\text{out}(x,y, t) \propto A \exp\left(i \frac{2\pi (f_0 + \beta t)}{v} x\right) \exp\left(-i \frac{2\pi \beta}{2v^2} x\right) $, which means if the input is a plane wave, the output will be deflected at a linearly varying angle of  $\theta(t) =  \arcsin(\frac{\lambda\,(f_0 + \beta t)}{v})$ and (cylindrically, such as in $x$ for a x-oriented AOD) focused down by a constant dioptric power of $ P_x =  \frac{\lambda\,\beta}{2v^2}$. This generates a lateral motion of the tweezer $x(t) = F \tan(\theta(t))\simeq \frac{\lambda F\,(f_0 + \beta t)}{v}$ as well as a meridional focal shift $\Delta F_x(t) = F^2 D \simeq \frac{\lambda F^2 \,\beta}{2 v^2}$.

\paragraph*{Multiple AODs}
Consider four AODs in two counter-propagating pairs, either directly stacked or $4f$ relayed as shown in our experimental setup. The AODs are namely: AOD-Ax with sound wave propagating in the $-x$ direction, AOD-Bx with sound wave propagating in the $+x$ direction, AOD-Ay with sound wave propagating in the $-y$ direction, AOD-By with sound wave propagating in the $+y$ direction. Then the final optical wavefront after experiencing all first-order diffractions will be decided by the following pupil function:
\begin{align}
     P(x,y, t)      &\propto  A_\text{Ax}(t+\frac{x}{v}) A_\text{Bx}(t-\frac{x}{v})  A_\text{Ay}(t+\frac{y}{v}) A_\text{By}(t-\frac{y}{v}) \cdot \nonumber \\
     & \quad\quad \exp\left({-i\left(\phi_\text{Ax}(t+\frac{x}{v}) + \phi_\text{Bx}(t-\frac{x}{v}) + \phi_\text{Ay}(t+\frac{y}{v}) + \phi_\text{By}(t-\frac{y}{v}) \right)}\right) \label{eq:Pxyt}\\
     & \hspace{0em} = \Tilde{A}(t) \cdot 
      \exp\Bigg( \underbrace{2\pi i\, \frac{ f_\text{Bx}(t)-f_\text{Ax}(t)}{v} x}_{\text{X-deflection}} 
     + \underbrace{2\pi i\, \frac{ f_\text{By}(t)-f_\text{Ay}(t)}{v} y}_{\text{Y-deflection}}  \notag  \\
& \qquad\qquad\quad  - \underbrace{2\pi  i\, \frac{\dot{f}_\text{Bx}(t)+\dot{f}_\text{Ax}(t) + \dot{f}_\text{By}(t)+\dot{f}_\text{Ay}(t)}{4 v^2} (x^2+y^2) }_{\substack{\text{spherical-lensing}}} \notag  \\
& \qquad\qquad\quad  - \underbrace{2\pi  i\, \frac{\dot{f}_\text{Bx}(t)+\dot{f}_\text{Ax}(t)-\dot{f}_\text{By}(t)-\dot{f}_\text{Ay}(t)}{4 v^2} (x^2-y^2)}_{\substack{\text{astigmatism}}}\Bigg),
\end{align}
where $\Tilde{A}(t) = A_\text{Ax}(t) e^{-i\phi_\text{Ax}(t) } \cdot  A_\text{Bx}(t) e^{-i\phi_\text{Bx}(t)} \cdot A_\text{Ay}(t) e^{-i\phi_\text{Ay}(t) } \cdot  A_\text{By}(t) e^{-i\phi_\text{By}(t)} $. One can conveniently set the constraint of  $\dot{f}_\text{Bx}(t)+\dot{f}_\text{Ax}(t)-\dot{f}_\text{By}(t)-\dot{f}_\text{Ay}(t) = 0$ to zero out astigmatism, while still having $(4-1=3)$ degrees of freedom to independently control the other three dimensions, namely: X-deflection, Y-deflection, and Z-spherical-lensing. This rationale underlies the device name of three-dimensional acousto-optic deflector lens (3D-AODL), despite the use of four AODs. In the final expression above, we only expand the wavefront amplitude modulation to the leading ($0^\text{th}$) order, and wavefront phase modulation to the $2^\text{nd}$ order. 


\subsubsection{Trap potential and force field}

After a Fourier transform using an ideal objective lens of focal length $F$, the wavefront on the back focal plane writes:

\begin{equation}
U_{\text{twz}}(X, Y, t) = \frac{e^{ikF}}{i \lambda F} \iint U_{\text{in}}(x, y) P(x,y,t) 
\, \exp\left[-i \frac{k}{ F}(x X + y Y)\right] \, dx\,dy
\end{equation}

where $k = 2\pi/ \lambda$.  We can also solve for the wavefront  on offset planes at offset $Z$ from the back focal plane to reconstruct the 3D wavefront amplitude:

\begin{align}
U_{\text{twz}}(X, Y, Z,t) &= \frac{e^{ik(F+Z)}}{i \lambda(F+Z)} \iint U_{\text{in}}(x, y) P(x,y,t) 
\exp\left( -i \frac{k}{2F}(x^2 + y^2) \right) \notag\\
& \qquad  \qquad  \qquad  \qquad  \qquad  \exp\left[ i \frac{k}{2(F+Z)} \left( (X - x)^2 + (Y - y)^2 \right) \right] dx\,dy \\
&\simeq \frac{e^{ik(F+Z)}}{i \lambda F} \,
\exp\left[ i \frac{k}{2F}(X^2 + Y^2) \right]\cdot \notag \\
&\iint U_{\text{in}}(x, y) P(x,y,t)  \,
\exp\left[ -i \frac{k}{ F}(x X + y Y) \right]
\exp\left[ -i \frac{k Z}{2 F^2}(x^2 + y^2) \right]
dx\,dy \label{Utwz_xyzt}
\end{align}

What follows is the optical dipole trapping potential, which is proportional to the intensity of the light field:

\begin{align}
V(X, Y, Z,t) & \propto  \left| U_{\text{twz}}(X, Y, Z,t)   \right|^2  
\end{align}

As well as its gradient, the optical dipole force  $\bm{F}=-\nabla |U_\text{twz}|^2$:
\begin{align}
\left\{
\begin{aligned}
F_x(X, Y, Z,t) & \propto  - \operatorname{Re} \left(U_{\text{twz}}^*\frac{\partial U_{\text{twz}}}{\partial X}   \right) \\
F_y(X, Y, Z,t) & \propto  - \operatorname{Re} \left(U_{\text{twz}}^*\frac{\partial U_{\text{twz}}}{\partial Y}   \right)  \\
F_z(X, Y, Z,t) & \propto  - \operatorname{Re} \left(U_{\text{twz}}^*\frac{\partial U_{\text{twz}}}{\partial Z}   \right)  
\end{aligned}
\right. \label{eq:force}
\end{align}

In the atom motion Monte-Carlo simulation below, we substitute Eq.~\ref{eq:Pxyt} into Eq.~\ref{Utwz_xyzt}, then substitute into Eq.~\ref{eq:force} to write down the equations of motion. 

\subsubsection{Monte-Carlo simulation of atom motion}

In the following, we use two sets of AOD and objective parameters to simulate atom trajectories in moving optical tweezers generated with the 3D-AODL and a microscope objective.

\textbf{(i)} A typical tweezer array apparatus  with medium-sized AOD active aperture of 7.5 mm (such as AAOptic DTSXY-400-80) and a moderate-NA objective of NA$=0.5$, an effective focal length of $F=6.5\,$mm, and an input aperture of $D_\text{obj}=7.5\,$mm. The 3D-AODL has the same aperture as the objective, so that a $1:1$ telescope is assumed to image the AOD aperture onto the objective aperture. We will call this \emph{the NA=0.5 scenario} for simplicity.

\textbf{(ii)} A state-of-the-art tweezer array apparatus with a large AOD active aperture of 15 mm (such as G\&H AODF 4085) and a high-NA objective of NA$=0.65$, an effective focal length of $F=6.5\,$mm, and an input aperture of $D_\text{obj}=11.1\,$mm. The 3D-AODL has a larger aperture than the objective, so that a $15:11$ telescope is assumed to de-magnify the AOD aperture onto the objective aperture. In the actual simulation, this is done by assuming the ``image AODL'' has an active aperture that's the same as the objective, and adopting an effective (slower) sound velocity by a factor of the magnification. We will call this \emph{the NA=0.65 scenario} for simplicity.  

We simulate the waveform on each AOD with $N_{\mathrm{pixel}}=100$ points that span $x\!\in[-D_\text{obj}/2, D_\text{obj}/2]$. 
We define the input beam onto all AODs as a cropped Gaussian with amplitude $A_0(x)\propto e^{-x^2/r^2}$ for $|x|\leq r$, with $r=D_\text{obj}/2$ , resulting in a filling factor of 1.

\begin{figure}
    \centering
    \includegraphics[width=0.9\linewidth]{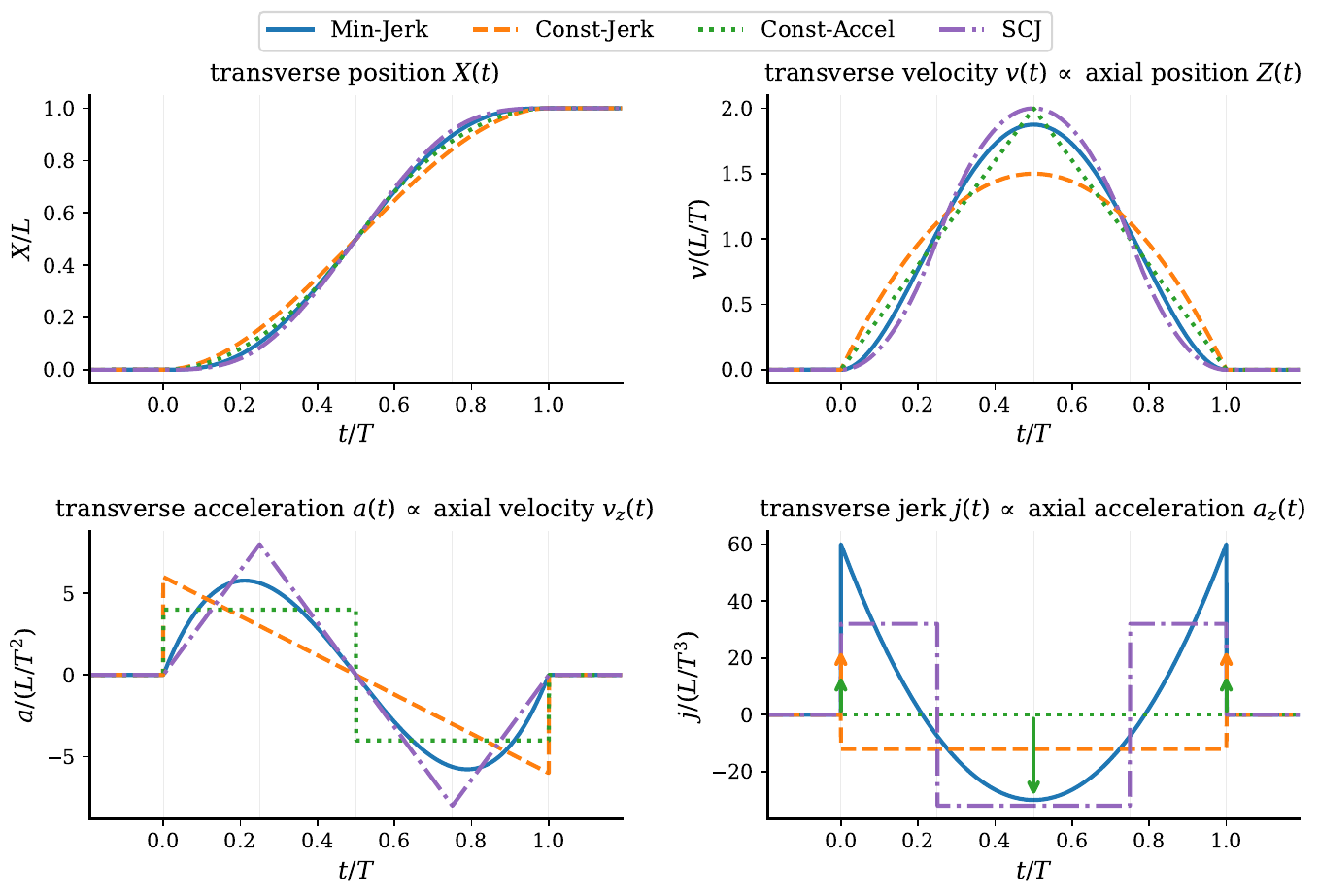}
    \caption{\textbf{Position, velocity, acceleration, and jerk} of four different kind of frequency ramps on a conventional AOD: minimum-jerk ramp, constant-jerk ramp, constant-acceleration ramp, and switching-constant-jerk ramp. The axial parameters here refer only to the focal quadrature that is addressed by the AOD. Arrows denote delta functions, representing a discontinuity in the temporal integral of the corresponding curve. }
    \label{fig:jerk}
\end{figure}

In both scenarios, we simulate the time-dependent tweezer potential and atom trajectories for either a $65\,\mu\text{m}$ $+X$ transport or a $65\sqrt{2}\,\mu\text{m}$ diagonal transport. We consider twenty drive configurations in each scenario, comprised of the product of four non-linear ramp functions plus a simple linear ramp, and four hardware configurations. As shown in Fig~\ref{fig:jerk}, the four non-linear ramp functions are:  minimum-jerk~(MJ) ramp, constant-jerk~(CJ) ramp, constant-acceleration~(CA) ramp, and switching-constant-jerk~(SCJ) ramp.
The four AODL hardware configurations are: ramps applied to a single AOD-Ax, to both AOD-Ax and AOD-Ay for diagonal motion, to AOD-Ax and AOD-Bx, or to all four AODs.

The minimum-jerk~(MJ) frequency ramp is defined as:
\begin{equation}
    f(t) = f_i + (f_f - f_i)\left[
    10\left(\frac{t}{T}\right)^3
    - 15\left(\frac{t}{T}\right)^4
    + 6\left(\frac{t}{T}\right)^5
    \right], \quad 0 \le t \le T.
\end{equation}

The constant-jerk~(CJ) frequency ramp (also known as the cubic ramp) is defined as:
\begin{equation}
    f(t) = f_i + (f_f - f_i)\left[
    3\left(\frac{t}{T}\right)^2
    - 2\left(\frac{t}{T}\right)^3
    \right], \quad 0 \le t \le T.
\end{equation}

The constant-acceleration~(CA) frequency ramp is defined as:
\begin{equation}
f(t)=
\begin{cases}
f_i + (f_f-f_i)\,2\left(\dfrac{t}{T}\right)^2, & 0\le t\le \dfrac{T}{2},\\[1.2ex]
f_i + (f_f-f_i)\left[1-2\left(1-\dfrac{t}{T}\right)^2\right], & \dfrac{T}{2}\le t\le T.
\end{cases}
\end{equation}

And the switching-constant-jerk~(SCJ) frequency ramp is defined as:
\begin{equation}
f(t)=
\begin{cases}
f_i + (f_f-f_i)\,\dfrac{16}{3}\left(\dfrac{t}{T}\right)^3,
& 0\le t\le \dfrac{T}{4},\\[1.2ex]
f_i + (f_f-f_i)\left[
-\dfrac{16}{3}\left(\dfrac{t}{T}-\dfrac{1}{2}\right)^3
+2\left(\dfrac{t}{T}\right)-\dfrac{1}{2}
\right],
& \dfrac{T}{4}\le t\le \dfrac{3T}{4},\\[1.2ex]
f_i + (f_f-f_i)\left[
\dfrac{16}{3}\left(\dfrac{t}{T}-1\right)^3 + 1
\right],
& \dfrac{3T}{4}\le t\le T.
\end{cases}
\end{equation}
where $T$ is the transport time. Both minimum-jerk~(MJ) and constant-jerk~(CJ) ramps are commonly considered for atom transport with conventional AODs~\cite{Bluvstein2022, 7r3w-8m61, Liu2019, Hwang25}. The MJ ramp minimizes the cumulative jerk-squared $J = \int_{0}^{T} \left( \frac{d^3x}{dt^3} \right)^2 dt$ over the entire trajectory for overall smoother motion \cite{Flash1688}, whereas the CJ ramp is a trajectory where the transverse acceleration of the tweezer evolves linearly in time. In addition, we consider two other ramp functions, the constant-acceleration~(CA) and switching-constant-jerk~(SCJ) ramps. The CA ramp is a two-piece function where the first(second) half undergoes constant transverse acceleration(deceleration); the SCJ ramp can be viewed as a middle-ground between the MJ and the CJ ramps, where the transverse acceleration follows piecewise linear evolution but never jumps. Notably, for conventional AOD-generated tweezer motion, the axial focal position follows the transverse velocity, and the axial focal velocity follows the transverse acceleration, and so on. This implies that when using conventional AODs, certain characteristics of the ramps matter more than with AODLs, such as the maximal velocity during the entire tweezer motion and the discontinuity in transverse acceleration.

\begin{figure}
    \centering
    \includegraphics[width=0.9\linewidth]{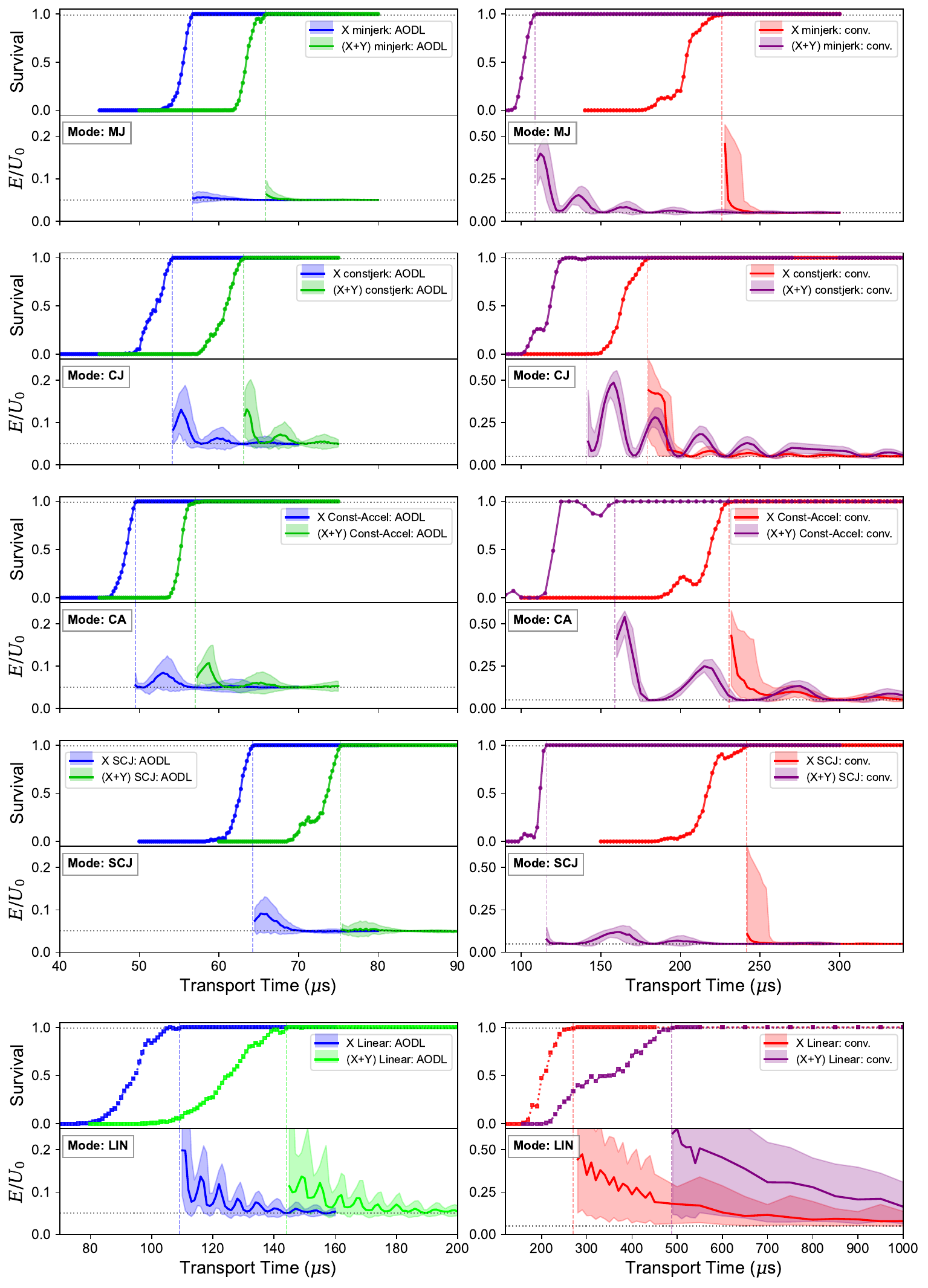}
    \caption{\textbf{Monte-Carlo simulation of atom transport using a NA=0.5 objective.}    (From top to bottom) Survival and final temperature using minimum-jerk~(MJ) trajectories; survival and final temperature using constant-jerk~(CJ) trajectories; survival and final temperature using constant-acceleration~(CA) trajectories; survival and final temperature using switching-constant-jerk~(SCJ) trajectories; survival and final temperature using linear ramp trajectories.  Transport distance is $65\,\mu\text{m}$ for the $X$ transport and $65\sqrt{2}\,\mu\text{m}$ for the diagonal transport.}
    \label{fig:NA=0.5}
\end{figure}

\begin{figure}
    \centering
    \includegraphics[width=0.9\linewidth]{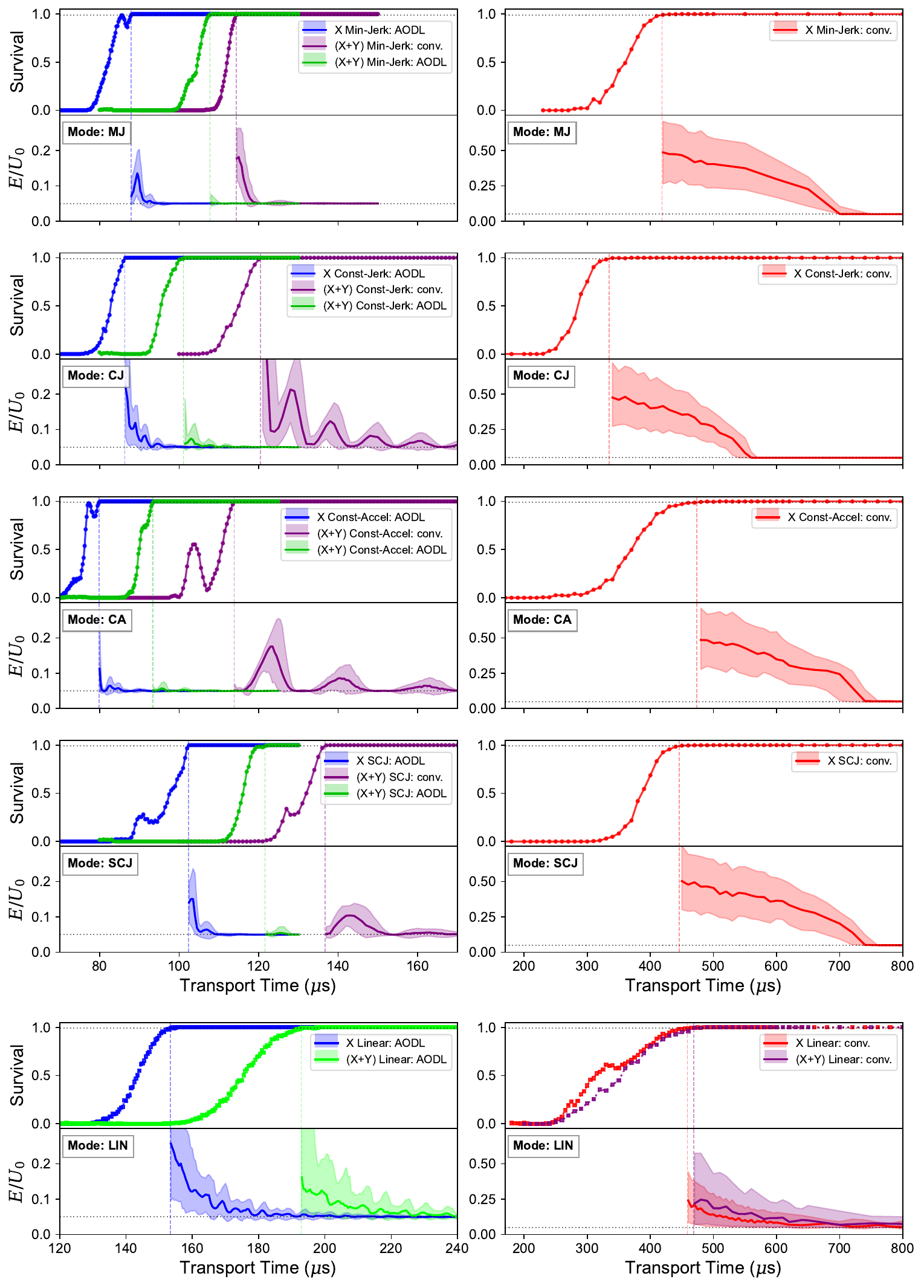}
    \caption{\textbf{Monte-Carlo simulation of atom transport using a NA=0.65 objective.}  (From top to bottom) Survival and final temperature using minimum-jerk~(MJ) trajectories; survival and final temperature using constant-jerk~(CJ) trajectories; survival and final temperature using constant-acceleration~(CA) trajectories; survival and final temperature using switching-constant-jerk~(SCJ) trajectories; survival and final temperature using linear ramp trajectories. Transport distance is $65\,\mu\text{m}$ for the $X$ transport and $65\sqrt{2}\,\mu\text{m}$ for the diagonal transport.}
    \label{fig:NA=0.65}
\end{figure}

We iteratively solve for atom motion in a time-evolving force field defined in Eq.~\ref{eq:force} using a fifth-order Runge–Kutta method for $N=1000$ atoms drawn from a random ensemble. We implement the Runge-Kutta ODE solver using SciPy's built-in RK45 method which assumes the accuracy of a fourth-order Runge-Kutta method while taking the steps of a fifth-order accurate formula to minimize truncation error and accurately simulate time-evolution. Initial positions and velocities are first sampled from Maxwell–Boltzmann distributions, and each atom’s total (kinetic + potential) initial energy $E_{\mathrm{init}}=K_{\mathrm{init}}+U_{\mathrm{init}}$ is then normalized to a fixed fraction (we use $1/20$ as a typical initial temperature) of the trap depth. This rescaling eliminates high-energy outliers while preserving the overall angular statistical characteristics of the ensemble.
Survival probability is defined as the fraction of atoms with final total energy $E_{\mathrm{final}}=K_{\mathrm{final}}+U_{\mathrm{final}}<0$ after transport. 
We present the Monte-Carlo results of $NA=0.5$ and $NA=0.65$ scenarios in  Fig~\ref{fig:NA=0.5} and~\ref{fig:NA=0.65}, respectively. The four AODL hardware configurations are color-coded as follows: ramps applied to a single AOD-Ax (red), to both AOD-Ax and AOD-Ay for diagonal motion (purple), to AOD-Ax and AOD-Bx (blue), or to all four AODs (green). In addition to the survival probability points, for points whose survival probability above 99\%, we also present the distribution of the final energy of the statistical ensemble, with the shaded region denoting the 68\% confidence interval.

When used on 3D-AODL, out of the four non-linear ramp functions, we observe the fastest transport with the constant-acceleration ramp. But if the objective is to minimize heating after multiple moves, then the minimum-jerk ramp shows the best suppression of motional heating, possibly due to the lack of rigid turns in the higher-order derivatives. Another fun observation is that compared to the other non-linear ramps, the constant-jerk ramp performs considerably better than other ramps when used on a single AOD, probably due to the lower maximum transverse velocity, resulting in lower maximum astigmatism.

In the NA=0.5 scenario, when using minimum-jerk frequency ramps, the simulation shows a 75\% transport time reduction for $+X$ transport and a  50\% transport time reduction for diagonal transport. This is because even with conventional 2D-AOD, diagonal transports are astigmatism-free (albeit out-of-plane). In contrast, when using linear frequency ramps, the conventional diagonal transport also fails miserably, due to the initial and final jump of the tweezer focus in the axial direction, as a result of the discontinuity of the first-order derivative of frequency.

In the NA=0.65 scenario, when using minimum-jerk frequency ramps, the simulation shows a 80\% (90\% if one also wants low final temperature) transport time reduction for $+X$ transport and a merely 6\% transport time reduction for diagonal transport. This is because at a higher NA of 0.65, the ratio of radial trap frequency to axial trap frequency reduces to a factor of three (compared to a factor of five in the NA=0.5 case), thus making the out-of-plane motion less problematic for atom survival. Notably, in both scenarios, the final temperature drops much faster as a function of transport time when using 3D-AODL -- a feature crucial for zone-based quantum computing, where multiple trips of long-range transport are performed, placing a tight requirement on the heating budget of each trip.

Additional parameters for all Monte-Carlo simulation runs are: objective aperture filling factor $=1$, laser wavelength $\lambda=0.808~\mu\mathrm{m}$,  and atom mass being $1.1419\times 10^{-25}$ kg, the mass of a $^{87}\mathrm{Rb}$ atom. Tweezer trap depth is set to be $E_{0}=h\times 20$ MHz for both NA scenarios (which implies that the NA=0.65 scenario needs roughly a factor of 2 less optical power per tweezer, compared to the NA=0.65 scenario), and the initial temperature is set to be $1/20$ of the trap depth.

\subsection{Waveforms without fading}
\subsubsection{General expression}

Because the first AODs have a narrower usable bandwidth than the second AODs, when generating multiple tweezers, it's natural to use single-tone waveforms on the first AODs (AOD-Ax and AOD-Ay), and to use multi-tone waveforms on the second AODs (AOD-Bx and AOD-By), as follows:
\begin{align}
\label{Eq:no-fading}
    \left\{
    \begin{aligned}
        V_\text{Ax}(t) &= A_\text{Ax}\cos\!\left(2\pi \int_0^{t} f_\text{Ax}(t')\,\mathrm{d}t' \right) \\
        V_\text{Ay}(t) &= A_\text{Ay}\cos\!\left(2\pi \int_0^{t} f_\text{Ay}(t')\,\mathrm{d}t' \right) \\
        V_\text{Bx}(t) &= \sum_{n=1}^{M_x} A_\text{Bx}^{(n)} 
        \cos\!\left(2\pi \int_0^{t} f_\text{Bx}^{(n)}(t')\,\mathrm{d}t' + \phi_\text{Bx}^{(n)} \right) \\
        V_\text{By}(t) &= \sum_{n=1}^{M_y} A_\text{By}^{(n)} 
        \cos\!\left(2\pi \int_0^{t} f_\text{By}^{(n)}(t')\,\mathrm{d}t' + \phi_\text{By}^{(n)} \right)
    \end{aligned}
    \right.
\end{align}
The frequency-position correspondence as a function of time can be written as follows: 
\begin{align}
    \left\{
    \begin{aligned}
        f_{Ax}(t) &= f_{0} - \frac{v}{2 \lambda F} \, X(t) + f_{Z}(t) \\
        f_{Ay}(t) &= f_{0} - \frac{v}{2 \lambda F} \, Y(t) + f_{Z}(t) \\
        f_{Bx}^{(n)}(t) &= f^{(n)}_{x0} + \frac{v}{2 \lambda F} \, X(t) + f_{Z}(t) \\
        f_{By}^{(n)}(t) &= f^{(n)}_{y0} + \frac{v}{2 \lambda F} \, Y(t) + f_{Z}(t) \\
    \end{aligned}
    \right.
\end{align}
Where $f_{Z}(t) = \frac{v^2}{2\lambda F^2} \int_0^{t} Z(t')\text{d}t'$. 

For an equi-spaced tweezer array, $f^{(n)}_{x0} = f_0 + \left(n - \frac{M_x + 1}{2}\right) \Delta f$ and $f^{(n)}_{y0} = f_0 + \left(n - \frac{M_y + 1}{2}\right) \Delta f$.


\subsubsection{Third  order intermodulation (IM3) and the Schroeder phase}
\label{Schroeder}

When sending multiple tones on an AOD, higher-order intermodulation alters the intensity of each tweezer and can lead to the generation of new tweezers at unwanted locations. This frequency mixing is a well-known effect, and is discussed in the supplementary material of \cite{Endres2016}. Briefly, on a single AOD, consider a frozen-time snapshot of the waveform composed of tones of equal amplitudes but different initial phases:
\begin{equation}
    V(x) = \sum_i \cos(k_i x + \phi^{(i)})
\end{equation}
Even though the acousto-optic response can be linear on the optical phase delay, the exponential functions of the optical field with respect to the optical phase introduce non-linearity beyond the weak-drive limit:
\begin{equation} 
    P(x) = \exp(i\, \Psi(x)) = \exp(i\, C \, V(x)) 
    =  1 + i\, C \, V(x) + (-1)\, C^2 \, V(x)^2 + (-i)\, C^3 \, V(x)^3 + \cdots
\end{equation}

Now the second-order terms contain differential and sum frequencies of the fundamentals:
\begin{equation} 
    \cos(k_i x + \phi^{(i)}) \cos(k_j x + \phi^{(j)})
    \rightarrow    
     \frac{1}{2}\cos\big((k_i+k_j) x + (\phi^{(i)}+ \phi^{(j)})\big) + \frac{1}{2}\cos\big((k_i-k_j) x + (\phi^{(i)}-\phi^{(j)})\big)
\end{equation}

While the second-order intermodulation (IM2) frequency is nominally outside of the AOD frequency resonance band, they can mix again with the fundamentals to form third (IM3) and higher order intermodulation tones, which produce frequencies within the AOD frequency resonance band that is either overlapping the fundamentals or at new locations. In order to minimize the influence of IM3 and higher-order intermodulation, it is useful to choose phases of the fundamentals, $\{\phi^{(i)}\}$, such that IM2 interferes destructively. Because without IM2, there's no IM3 or higher-order intermodulation.  

In the special case of $M$ equi-spaced frequency tones, we use the so-called Schroeder phase\cite{Schroeder1970, Friese1997}, which eliminates the nearest neighbor IM2 by placing them uniformly around a circle of the complex plane, and almost eliminates the second-nearest neighbor IM2 by occupying all-but-one along the circle, and so on... 

\begin{equation} 
    \phi^{(i)} = 2 \pi \times\frac{i(i-1)}{2 (M-1)}  
    \quad \longleftrightarrow    \quad
     \phi^{(i+1)} - \phi^{(i)} = 2 \pi \times\frac{i}{M-1} 
\end{equation}


\subsection{Fading-Shepard Waveforms}

\begin{figure}
    \centering
    \includegraphics[width=1\linewidth]{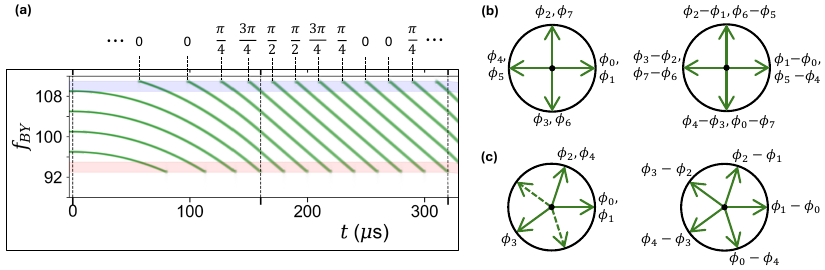}
    \caption{\textbf{Schroeder phase for fading-Shepard waveforms.}
    (a) \textit{Schroeder phase assignment} for $M=4$ fading-Shepard waveform used in Fig.~4 in the main manuscript. 
    (b) \textit{Phase assignment for an even number of tones,} using $M=4$ as an example.  The left panel displays the complex-valued fundamental tones, while the right panel shows the complex-valued second-order intermodulation product (IM2) resulting from nearest-neighbor frequency differences. The phases of the fundamentals have periodicity $(2M)$ if $M$ is even. 
    (c) \textit{Phase assignment for an odd number of tones,} using M=5 as an example. The phases have periodicity $M$ when $M$ is odd. 
    }
    \label{fig:Shepard-Schoeder}
\end{figure}

\subsubsection{General expression}
The general expression for the fading-Shepard waveforms can be viewed as a modified version of the no-fading waveforms in Eq.~\ref{Eq:no-fading}, with amplitude of each tone fading in and out in time:

\begin{align}
\label{Eq:fading-Shepard}
    \left\{
    \begin{aligned}
        V_\text{Ax}(t) &= \sum_{n=-\infty}^{\infty} A_\text{Ax}^{(n)}(t) \cos\!\left(2\pi \int_0^{t} f_\text{Ax}^{(n)}(t')\,\mathrm{d}t' \right) \\
        V_\text{Ay}(t) &=  \sum_{n=-\infty}^{\infty} A_\text{Ay}^{(n)}(t) \cos\!\left(2\pi \int_0^{t} f_\text{Ay}^{(n)}(t')\,\mathrm{d}t' \right) \\
        V_\text{Bx}(t) &= \sum_{n=-\infty}^{\infty} A_\text{Bx}^{(n)}(t) 
        \cos\!\left(2\pi \int_0^{t} f_\text{Bx}^{(n)}(t')\,\mathrm{d}t' + \phi_\text{Bx}^{(n)} \right) \\
        V_\text{By}(t) &= \sum_{n=-\infty}^{\infty} A_\text{By}^{(n)}(t) 
        \cos\!\left(2\pi \int_0^{t} f_\text{By}^{(n)}(t')\,\mathrm{d}t' + \phi_\text{By}^{(n)} \right)
    \end{aligned}
    \right.
\end{align}

Where the frequency evolution of each tone can be written as follows: 
\begin{align}
    \left\{
    \begin{aligned}
        f_{Ax}^{(n)}(t) &= f_{0} - \frac{v}{2 \lambda F} \, X(t) + f_{x, Z}^{(n)}(t) \\
        f_{Ay}^{(n)}(t) &= f_{0} - \frac{v}{2 \lambda F} \, Y(t) +  f_{y, Z}^{(n)}(t) \\
        f_{Bx}^{(n)}(t) &= f_{0} + \frac{v}{2 \lambda F} \, X(t) +  f_{x, Z}^{(n)}(t) \\
        f_{By}^{(n)}(t) &= f_{0} + \frac{v}{2 \lambda F} \, Y(t) +  f_{y, Z}^{(n)}(t) \\
    \end{aligned}
    \right.
\end{align}
where $ f^{(n)}_{\mu, Z}(t) = \frac{v^2}{2\lambda F^2} \int_0^{t} Z(\tau)\text{d}\tau  + \left(n+ \xi_{\mu} \right) \Delta f $,  $\, \mu \in \{x, y\}$.

\vspace{5pt}
The amplitude evolution of each tone on the first AODs, AOD-Ax and AOD-Ay, writes:

\begin{align}
\label{Eq:intensity1}
A^{(n)}_{Ax} &=
\begin{cases}
1,
& \left|f^{(n)}_{Ax,Z}\right| \le \dfrac{(1-\eta)\,\Delta f}{2},
\\[6pt]
0,
& \left|f^{(n)}_{Ax,Z}\right| \ge \dfrac{(1+\eta)\,\Delta f}{2},
\\[6pt]
\cos^{\!\;p_{Ax}}\!\left[
\dfrac{\pi}{2\eta}
\left(
\dfrac{\left|f^{(n)}_{Ax,Z}\right|}{\Delta f}
- \dfrac{1}{2}
\right)
+ \dfrac{\pi}{4}
\right],
& \text{otherwise}.
\end{cases}
\\[6pt]
A^{(n)}_{Ay} &\text{ is defined identically with } x \to y. \nonumber
\end{align}

and the amplitude evolution of each tone on the second AODs, AOD-Bx and AOD-By, writes:

\begin{align}
\label{Eq:intensity2}
A^{(n)}_{Bx} &=
\begin{cases}
1,
& \left|f^{(n)}_{Bx,Z}\right| \le \dfrac{(M_x -\eta)\,\Delta f}{2},
\\[6pt]
0,
& \left|f^{(n)}_{Bx,Z}\right| \ge \dfrac{(M_x +\eta)\,\Delta f}{2},
\\[6pt]
\cos^{\!\;p_{Bx}}\!\left[
\dfrac{\pi}{2\eta}
\left(
\dfrac{\left|f^{(n)}_{Bx,Z}\right|}{\Delta f}
- \dfrac{M_x}{2}
\right)
+ \dfrac{\pi}{4}
\right],
& \text{otherwise}.
\end{cases}
\\[6pt]
A^{(n)}_{By} &\text{ is defined identically with } x \to y. \nonumber
\end{align}
where $\eta$ is the fading duty, defined as the spectral width of the fading zone divided by the frequency spacing, and is set to $\eta=1/2$ in the main manuscript. 
The fading orders need to satisfy $p_{Ax}+p_{Bx}=p_{Ay}+p_{By}=1$, ensuring that the sum of \emph{old} and \emph{new} tweezer powers remains constant. 
As shown in Tab.~II in the main manuscript, when generating fading-Shepard waveforms for a single tweezer, we set $p_{Ax}=p_{Bx}=p_{Ay}=p_{By}=0.5$, and thus the fading powers are the same on the first AODs and the second AODs. However, when generating fading-Shepard waveforms for an array of tweezers, we set $p_{Ax}=p_{Ay}=1$ and $p_{Bx}=p_{By}=0$, so that the intensity of tweezers in the bulk of the array stays constant. This is also shown in the fading of intensities in the spectrograms in Fig.~3 VS Fig.~4 in the main manuscript.

As for the relative phases of the tones on AOD-Bx and -By, we adopt a generalized version of the Schroeder phase (as shown in Fig.~\ref{fig:Shepard-Schoeder}) similar to the no-fading Schroeder phases: 

\begin{align}
\label{Eq:Shepard-Schoeder}
    \left\{
    \begin{aligned}
        \phi_{\mathrm{Bx}}^{(n)} &=
            \operatorname{mod}\!\left(
            2\pi\,\frac{n(n-1)}{2M_x},
            \,2\pi
            \right) \\
        \phi_{\mathrm{By}}^{(n)} &=
            \operatorname{mod}\!\left(
            2\pi\,\frac{n(n-1)}{2M_y},
            \,2\pi
            \right)
    \end{aligned}
    \right.
\end{align}

We use $2M$ instead of $2(M-1)$ in the denominator because there are actually $M+1$ fundamental tones during fading. A simple check will reveal that the infinite series $\phi^{(n)}$ has a periodicity of $M$ if $M$ is odd, and a periodicity of $2M$ if $M$ is even.

\subsubsection{Interlaced VS simultaneous fading}

\begin{figure}
    \centering
    \includegraphics[width=1\linewidth]{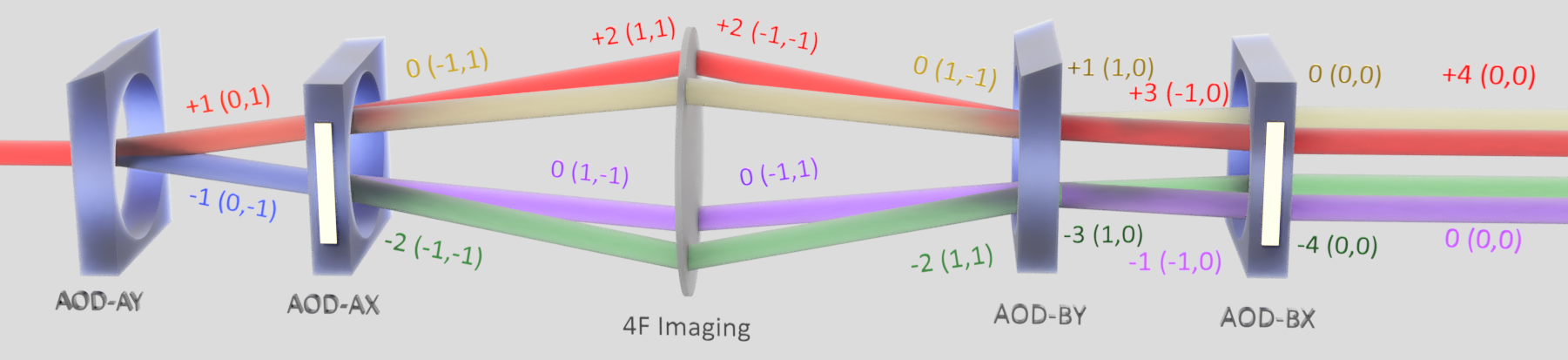}
    \caption{\textbf{Visualization of contributors to a single spatially overlapped tweezer during simultaneous fading.} Each relevant light path is colored differently for distinguishability, and labeled by its temporal and spatial frequencies in the format of ``$\Delta f \,(k_x, k_y)$''. Note that the yellow and purple paths have the same temporal and spatial frequencies in the end, leading to static constructive/destructive interference. 
    }
    \label{fig:simultaneous}
\end{figure}

The temporal location of the fading zones of the AOD pair in x (y) is dependent on the frequency offsets $\xi_x$ ($\xi_y$). For example, if we set $\xi_x = \xi_y$, the fading of the x-AOD pair and the y-AOD pair happen simultaneously. 

During simultaneous fading of single tweezer generation, each AOD is driven with two frequency tones and splits the input light two ways, as shown in Fig.~\ref{fig:simultaneous}. After all four AODs, there should be a total of 16 output light rays, but only four of them are pointing at the correct spatial location and thus contribute to the generation of the center (useful) tweezer, while the others form shadow tweezers (Sec.~\ref{shadow}). The four constituents of the center tweezer are shown in red, yellow, purple, and green on the right-hand side of Fig.~\ref{fig:simultaneous}, and two of them (yellow and purple) have the same optical frequency while the other two have different and unique optical frequencies. When two light rays of the same optical frequency overlap in spatial mode, the superposition is akin to a static Mach–Zehnder interferometer, with the total intensity depending on the path-length difference between the two (constructive vs. destructive interference). In a real-life optical setup, the path-length difference is easily susceptible to mechanical instability (such as temperature changes or vibrations), which makes simultaneous fading suboptimal. 

However, if we set $\xi_x$ and $\xi_y$ offset from each other by $1/2$, while also setting $\eta \leq 1/2$, the fading zones of the x-AOD pair and the y-AOD pair interlace each other and never overlap. In the interlaced fading regime, the two constituents of the center tweezer always have different optical frequencies, separated by a few MHz. When two light rays of different optical frequencies overlap in spatial mode, the total optical field intensity oscillates at a beatnote frequency equal to the difference between the two. This MHz scale beatnote frequency is much higher than the typical trap frequency of an atom in an optical tweezer, so as far as the atom is concerned, the constituents simply sum up in intensity instead of amplitude. For this reason, we choose interlaced fading for all the example tweezer trajectories in the main manuscript. 

An exception in which simultaneous fading results in simple intensity summation of the constituents occurs when the array geometry is rectangular rather than square, i.e., $\Delta f_x \neq \Delta f_y$.

\section{Experimental Methods}
\subsection{Experiment apparatus}

\begin{figure}
  \includegraphics[width=\linewidth]{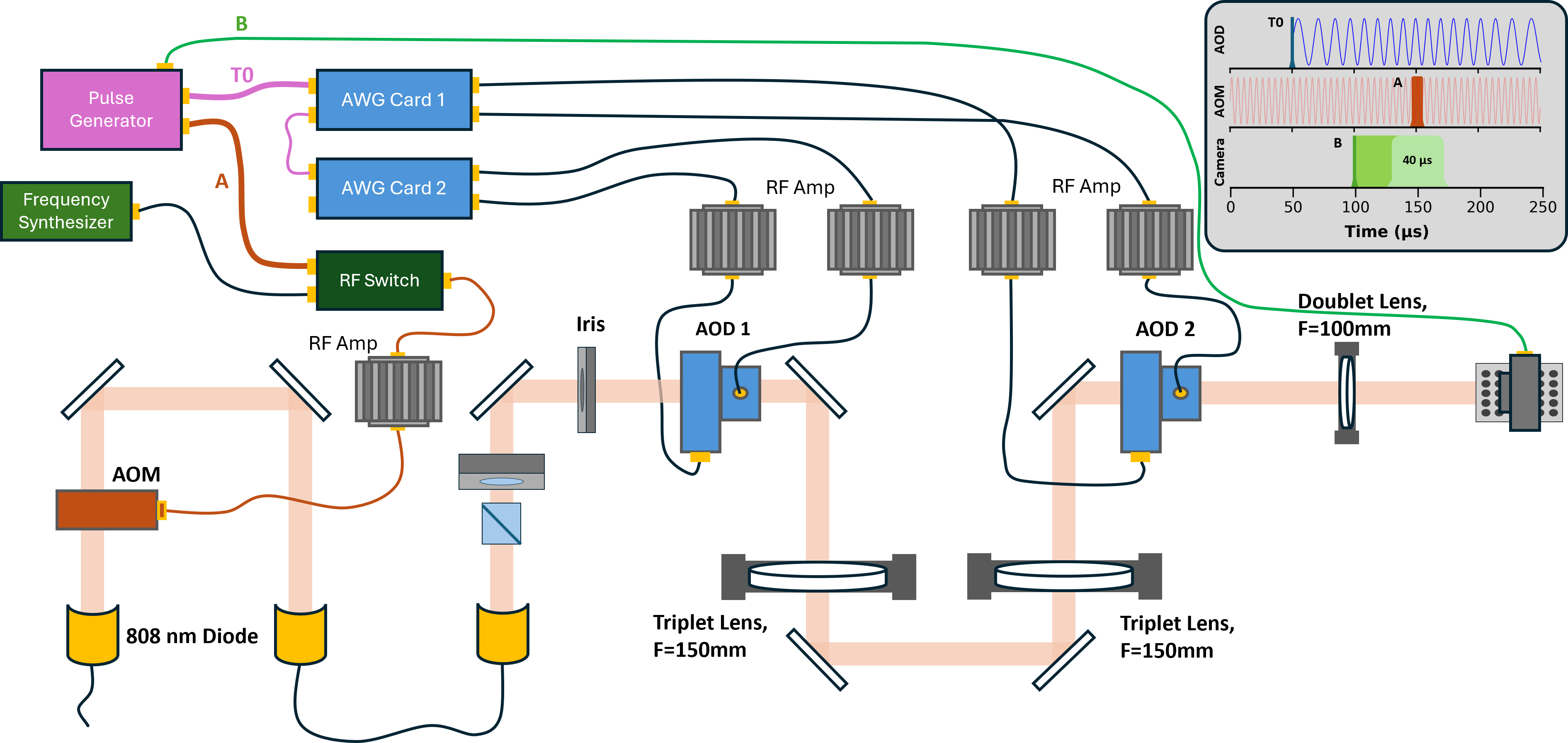}

  \vspace{5pt}

    \justifying
    \captionof{figure}{\textbf{Optical setup and RF connection diagram.}
    The laser output (808 nm) is modulated with an IntraAction AOM driven by a program- mable trigger delay `A'.
    The start of each trigger pulse `A' is heralded by pulse `T0' which initiates AWG cards and respective
    waveforms into X and Y channels of AOD~1 (AOD-A) and~2 (B).
    The modulated laser output is fiber coupled and sent through a 4-F imaging system consisting of two
    Hasting triplet lenses of $F=150\,\mathrm{mm}$.
    Imaging is performed with a doublet lens of $F=100\,\mathrm{mm}$ onto a camera mounted on a motorized translation stage.}
    \label{fig:opticalsetup}
\end{figure}

Our experimental apparatus consists of two orthogonally oriented 2D acousto-optic deflectors (AODs; DTSXY-400-800.860 and DTSXY-400-780-002, AA Optoelectronics) whose active apertures are imaged onto each other with lateral inversion using a 4-\textit{f} imaging system, as shown in Fig.~\ref{fig:opticalsetup}. The 4f-relay comprises two Hastings triplets with \textit{f} = 150 mm mounted on stages providing two rotational and two translational degrees of freedom. Hastings triplets were chosen to balance tunability and compactness while minimizing aberrations under small misalignments; similar configurations have been reported previously~\cite{Rozsa2016,Kirkby2010}.

The AOD response time is $\tau = L/v = 11.54,\mu\text{s}$, where $L = 7.5$ mm is the active aperture and $v = 650$ m/s is the acoustic velocity in the crystal. To ensure precise overlap of the active apertures, the second AOD is mounted on a translation stage for fine lateral adjustment. An 808-nm diode laser (Thorlabs LD808-SEV500) is shuttered by an acousto-optic modulator (IntraAction ATM-2701A2) for stroboscopic imaging, fiber-coupled, and sent through the 3D-AODL. The input light is vertically polarized using a polarizing beam splitter, with a half-wave plate used to optimize diffraction efficiency on the first AOD. Both AODs are driven by two synchronized dual-channel arbitrary waveform generators (Spectrum M4i.6631-x8). The resulting optical tweezers are imaged using a 100-mm doublet onto a CMOS camera (Thorlabs Zelux) mounted on a motorized translation stage (Thorlabs MTS50-Z8).

\paragraph{\textbf{Unit Conversion}}
Spatial coordinates are reported in converted units, $\mu\text{m}^*$, to reflect typical atom trapping scales. We chose a focal lens of $F = 100$ mm instead of an objective in order to maximize the image size of tweezers on our CCD camera and obtain better metrics for beam quality. In most of the main manuscript, this effective focal length is rescaled to an effective $F^*=6.5$~mm objective (as in the Monte Carlo simulation). This corresponds to multiplying the lateral positions (pixel index $\times$ $3.45\,\mu$m) by $0.065$ and the axial positions (translation stage) by $0.065^2 = 4.23\times10^{-3}$. Using converted units, the static tweezer has a waist radius of 1.1(1) $\mu\text{m}^*$ and a Rayleigh range of 4.2(3) $\mu\text{m}^*$, and $1\,$MHz of frequency difference maps to 8.125 $\mu\text{m}^*$ spacing.


\subsection{Automated data acquisition}

Our experimental data for each run consists of a four-dimensional array (dimensions being $X, Y, Z, t$) with each point storing the intensity captured on the CCD camera. Timesteps in T were taken with a 500 ns shutter time using the SRS delay generator. In comparison to our tweezer movement times (Ranging from 50 to 200 $\mu$s spanning 90 to 100 MHz), this shutter time was minimal and showed no significant blurring per frame. Data acquisition loops (Illustrated in Algorithm 1) initialized a translation stage (z=0.0 mm) to move the CCD camera and take images at designated times T. After imaging all times T, the translation stage was moved in z and acquisition was repeated until the maximal range of our translation was reached (z=25.0mm).

\begin{algorithm}
\caption{Data Acquisition}
\KwData{Delay/Pulse Generator, Waveform Generator Channels (Ax,Bx,Ay,By), Camera, Translation Stage}
\KwResult{Four-Dimensional array (X,Y,Z,T) with Three-Dimensional Tweezer Profiles across timesteps.}
Initialize connections to all devices. Move translation stage to z-offset $0.00 \text{ mm}$.

\textbf{Define} sets of times $\{t\}$ and z-positions $\{z\}$ for data acquisition points.

\For{$z \in \{z_{camera}\}$}{
    Move translation stage to $z$\;
    \For{$\tau \in \{t_{delay}\}$}{
        Set pulse generator delay to $\tau$. Pulse on gates AOM as seen in Fig.~\ref{fig:opticalsetup}\;
        Trigger Waveform Generators using pulse generator T0 output.\;
        Take $X$ images with CCD camera\;
        Store images in designated folder $(z,\tau)$\;
    }
}
\end{algorithm}

Each X and Y channel of AOD 1 and AOD 2 (Fig.~\ref{fig:opticalsetup}) were driven with independently programmable channels of an arbitrary waveform generator (Spectrum Instrumentation M4i.6631-x8). The waveform generator cards were synced to each other using internal triggers while an external trigger T0 was used to initialize waveform runs. Imaging was performed with a Thorlabs CCD camera attached to a translation stage with the translation axis along the direction of propagation of the tweezers. As the CCD camera lacked high speed shuttering capabilities, the input laser was shuttered instead to only provide light at desired times of imaging. Camera exposure time was synced to be a multiple of the waveform time span to ensure that each image took the sum of \textit{N} shuttered pulses. Shuttering was performed using an AOM (IntraAction) that was driven with a constant 200 MHz tone switched on at desired times upon trigger by a delay generator.

\section{Design Caveats}

\subsection{Limited first AOD bandwidth }
\label{bandwidth}

\begin{figure}
    \centering
    \includegraphics[width=1\linewidth]{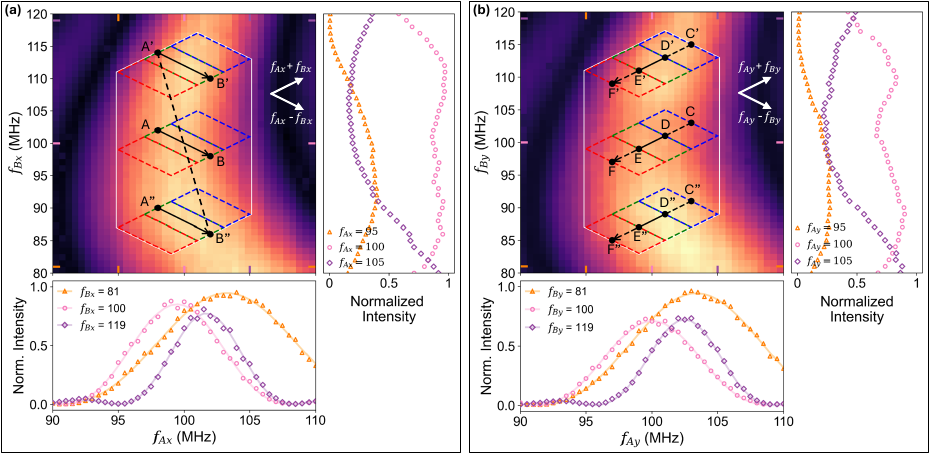}
    \caption{  \textbf{(a) Overall diffraction efficiency of the x-AOD pair.}  
     \textbf{(b) Same for the y-AOD pair.} Red/blue/green dashed rectangles label the lower/upper/non-fading zones for a single tweezer, while a 3-stack of them labels three different tweezers. The white polygon cover of these rectangles labels the usable zone of the AODL.
     }
    \label{fig:intensity}
\end{figure}

As stated previously,  the 3D-AODL enjoys the TPM-broadened bandwidth on the drive frequency of the second AOD but not that of the first, because the acoustic frequency of the first AOD alters the incident angle onto the second AOD, making the usable bandwidth of the first AODs (AOD-Ax and AOD-Ay) a few times narrower than the usable bandwidth of the second AODs (AOD-Bx and AOD-By). 

As shown in Fig.~\ref{fig:intensity}, astigmatism-free tweezer motion in the $X$ direction will follow lines on which $f_{Ax}+f_{Bx}$ is kept constant, such as $\overline{AB}$, $\overline{A'B'}$ and $\overline{A''B''}$ in the left panel. Due to the limited bandwidth of $f_{Ax}$, these astigmatism-free trajectories do not go over the full span of the bandwidth of the second AOD. However, if one must use the full span of the second AOD to reach positions far from the center, they can sacrifice some astigmatism-free-ness and go along a trajectory that is not fully astigmatism-free, such as $\overline{A'B''}$. 

In both panels of Fig.~\ref{fig:intensity}, the dashed blue(red) rectangle is the upper(lower) fading zone of a single tweezer, and the dashed green region in between is the non-fading zone. Astigmatism-free tweezer motion in the $Z$ direction will follow lines on which $f_{Ax}-f_{Bx}$ and $f_{Ay}-f_{By}$ is kept constant, such as $\overline{CF}$, $\overline{C'F'}$ and $\overline{C''F''}$ in the right panel. Note that each line is split into three segments, which are in the upper fading zone, the non-fading zone, and the lower fading zone, respectively. 

It's also worth noting that although the astigmatism-free motion range in the $X-Y$ plane is limited by the first AOD bandwidth, the span of the tweezer array does not share the same limit, because we use multiple tones on the second AODs to generate the $M_x$-by-$M_y$ array. In other words, while each individual tweezer must limit its range of motion within the red-green-blue dashed rectangular area (if the aim is astigmatism-free motion), the array can still span the cover of all the individual red-green-blue rectangles, as indicated by the white polygon.


\subsection{Sensitivity to optical misalignment}

\begin{figure}
    \centering
    \includegraphics[width=1\linewidth]{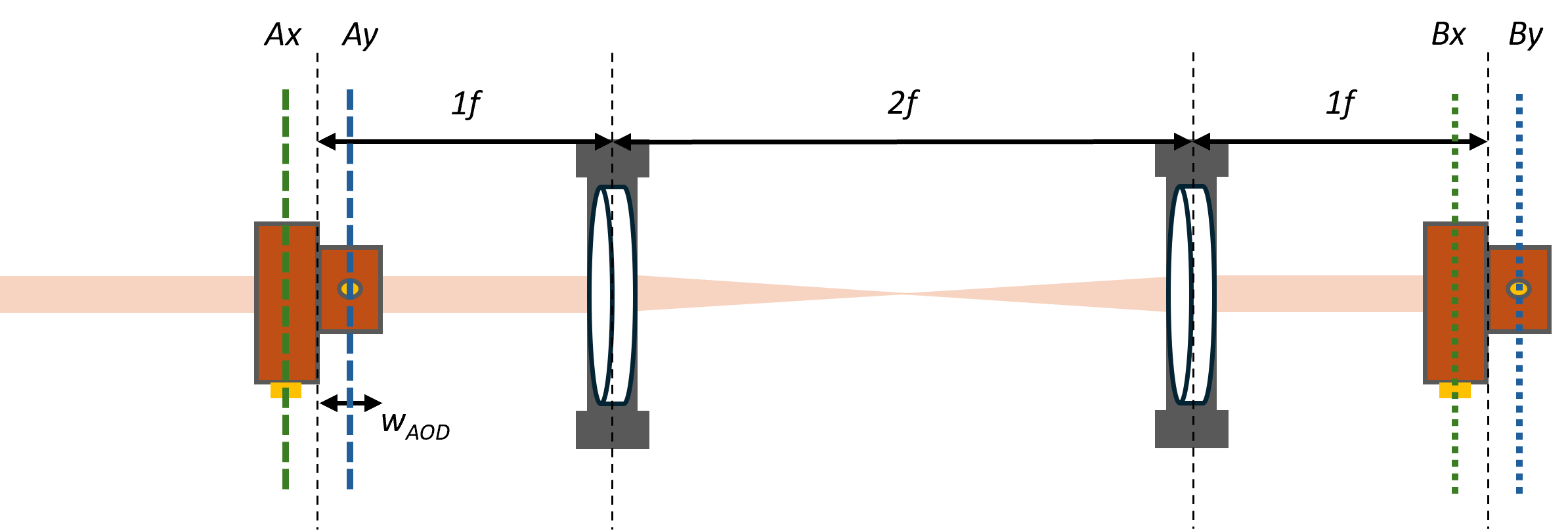}
    \caption{ The optical path here is shown with no misalignment as a demonstration. Even with perfect alignment and 4F relaying between the two crossed-AODs, observe that magnification mismatch will occur because the x- and y-AOD pairs are not exactly located on the 4F imaging planes. }
    \label{fig:magnificationimage}
\end{figure}

The quality of optical tweezers generated with the AODL was highly sensitive to the alignment of optical components. Small defects in the angle and spatial calibration of the triplet lenses used for 4-\textit{f} imaging enhanced higher order optical aberrations. To obtain good calibration of our optical path elements, a beam profiler was used to diagnose the beam shape before and after the addition of each new path element. For example, before placing the first lens of the 4-\textit{f} imaging system, a beam profiler was placed two focal lengths away from AOD 1. The spatial calibration of the first triplet lens was then performed to place the focused beam center at the location found using the beam profiler. As proper operation of the 3D-AODL relies on the superposition of all four acoustic wavefronts through 4F imaging we required additional alignment procedures to account for offsets or tilts introduced on the superimposed planes. 

As the 2D-AODs comprising our 3D-AODL have a finite width, it is impossible to overlay all four acoustic wavefronts spatially. Additionally, as the piezoelectric crystals inside the AOD itself have differing positions the acoustic wavefront for each AOD will also begin at different locations. Effectively, this means that each AOD waveform will experience a delay time mismatch of
\begin{equation} 
    P_i(x,y, t)  \simeq  \big( 1 + i C \, V_i(t-\frac{x-x_\text{err}}{v})\big)
\end{equation}

    
This misalignment can be observed by measuring the intensity of tweezer during fading. 
To zero-out the difference in acoustic delay of each counterpropagating AOD pair, we perform intensity pulse modulation on the drive waveform of said pair, and maximize the temporal overlap of the acoustic waves passing through the optic aperture by maximizing the pulsed tweezer intensity. 
After determining the delay times $x_\text{err}$ for each AOD $i$, we may compensate for the delay time mismatch by adding a delay time to each waveform. This calibration returns the desired phase wavefront of 
\begin{equation} 
    P_i(x,y, t)  \simeq  \big( 1 + i C \, V_i(t-\frac{x}{v})\big)
\end{equation}


Consider the issue of 4F-magnification once again, as shown in Fig.~\ref{fig:magnificationimage}. As our AODs have a finite width, it is difficult to ensure that the Ax, Bx and Ay, By AOD pairs will be imaged onto each other at both equivalent spatial locations and magnifications. As such, we experience a persistent magnification error using the 4F imaging despite physical alignment. As the magnification mismatch  is physically equivalent to a rescaling of the acoustic velocity on AODs Bx and By, we compensate for the mismatched image sizes by adjusting our frequency ranges by an appropriate dilation factor. For example, in the ideal case with no magnification mismatch the frequency range of AODs Ax, Ay, Bx, and By would be $(90,110)$ MHz. However, using these parameters causes tweezers that are fading in and out to mismatch spatially due to mismatched edge frequencies failing to combine properly. By adding a dilation factor such that the frequency range of AODs Ax, Ay $\in (90,110)$ MHz and the range of AODs Bx, By $\in (90.2,109.8)$ MHz. As our AOD width is typically small compared to the length of the 4F imaging system, the dilation factor is likewise small in magnitude. 

We emphasize that the alignment and compensation procedures described were performed manually under typical laboratory conditions, relying on iterative adjustments of optical elements. As such, the residual trap position and intensity modulations observed (as shown in Fig.~4 in the main manuscript) should not be regarded as intrinsic limitations of the device. Rather, we expect that industrial-grade alignment tools and automated calibration methods could substantially reduce these imperfections, further improving spatial overlap and uniformity of the generated tweezers.

\subsection{Shadow tweezers}
\label{shadow}

During tweezer array movements involving fading-Shepard waveforms, shadow tweezers arise during times where tones fade in and out. These shadow tweezers are the result of sequential deflections that land outside the grid of tweezer array. Consider the case of a single tweezer held at a constant Z-offset using fading-Shepard tones. For a single tone, we set $p_{Ax}=p_{Bx}=p_{Ay}=p_{By}=0.5$. 




Observe in Eq.~\ref{Eq:intensity1} and \ref{Eq:intensity2}, in the fading regions $\frac{(1-\eta)\Delta f}{2} < |f| < \frac{(1+\eta)\Delta f}{2}$, the amplitude of the tones is no longer unity. Specifically, in the fading region, we have both the fade-in and fade-out tones with a frequency separation of $f^\text{upper} - f^\text{lower} = \Delta f$, while the amplitudes follows:

\begin{align}
    \left\{
    \begin{aligned}
    A^\text{upper} &=  \cos^{\!\;0.5}\!\left[
\dfrac{\pi}{2\eta}
\left(
\dfrac{f^{\text{upper}}}{\Delta f}
- \dfrac{1}{2}
\right)
+ \dfrac{\pi}{4}
\right] \\
    A^\text{lower} &=  \cos^{\!\;0.5}\!\left[
\dfrac{\pi}{2\eta}
\left(
\dfrac{-f^{\text{lower}}}{\Delta f}
- \dfrac{1}{2}
\right)
+ \dfrac{\pi}{4}
\right] 
    \end{aligned}
    \right.
\end{align}

For example, take the instance of interlaced fading, when on the x-AOD pair the fade-in tone has faded exactly halfway into the waveform, both AOD-Ax and AOD-Bx are driven with two tones of $f^\text{upper} = f_0 +\frac{\Delta f}{2}$ and  $f^\text{lower} = f_0 -\frac{\Delta f}{2}$, and both amplitudes $A^\text{upper}=A^\text{lower}=(0.5)^\frac{1}{4}$, respectively. The tweezer generated by design is the intensity sum of  (upper tone on AOD-Ax) $\times$ (upper tone on AOD-Bx) and (lower tone on AOD-Ax) $\times$ (lower tone on AOD-Bx), which are mode-matched spatially. The shadow tweezers are the byproduct of (upper tone on AOD-Ax) $\times$ (lower tone on AOD-Bx), and (lower tone on AOD-Ax) $\times$ (upper tone on AOD-Bx), located on either side of the target tweezer with an x-offset of $\pm \frac{2 \lambda F }{v} \Delta f$.

The amplitude function for a single tweezer is designed to keep constant intensity of the target tweezer, so during fading of the x-AOD pair, two shadow tweezers emerge briefly with $\pm$ offset in $X$, and during fading of the y-AOD pair, two shadow tweezers emerge briefly with $\pm$ offset in $Y$. Similarly, in the case of a tweezer array, shadow tweezers emerge around the edges of the target array. These shadow tweezers form a $(M_x+2)$-by-$M_y$ extended grid pattern during x-AOD fading, and a $M_x$-by-$(M_y+2)$ extended grid pattern during y-AOD fading. 

When using a 3D-AODL-generated moving tweezer array to selectively pick up atoms from a static atom array, one needs to pay special attention to avoid picking up atoms at the shadow tweezer locations. If the locations of any shadow tweezers overlap with the location of the static array, the motion should start and end during the non-fading zones, when shadow tweezers are not present. 


\subsection{The acoustic-irising effect}
\label{acoustic-irising}

\begin{figure}
    \centering
    \includegraphics[width=1\linewidth]{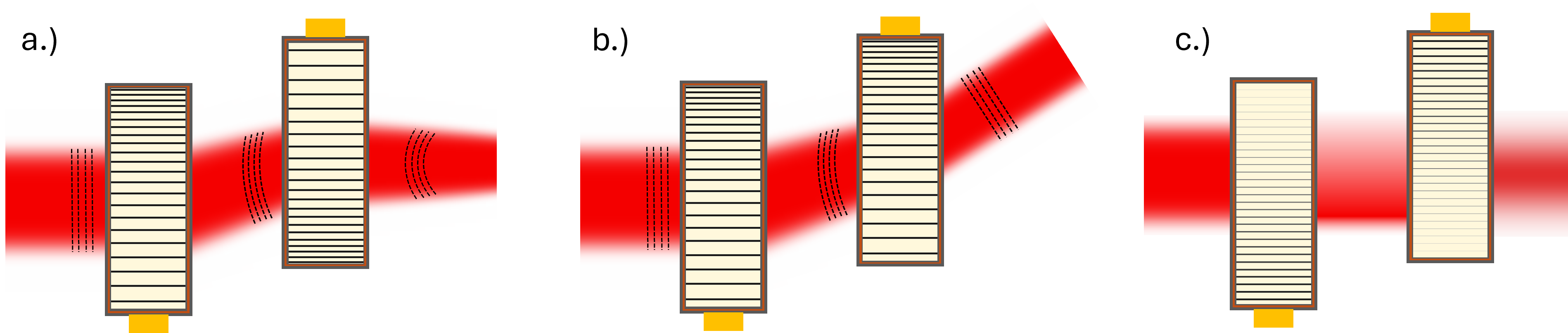}
    \caption{\textbf{(a,b): acoustic deflection and acoustic lensing; (c): acoustic irising.}}
    \label{fig: acoustic irising}
\end{figure}

As shown in Fig.~\ref{fig: acoustic irising}, in addition to acoustic deflection and lensing, the second-order amplitude modulation of the optical wavefront introduces optical irising. This is due to the quadratic amplitude term seen in Eq.~\ref{s5}, which modulates the amplitude of an incoming waveform. Under a drive waveform $V(t) = A(t) \cos (2\pi f_0 t)$ with varying amplitude and constant frequency, the first-order diffraction wavefront will exhibit the term $\frac{-A^{(2)}(t)}{2!v^2}x^2$. As shown in the pair of AODs in \ref{fig: acoustic irising}(c), both AODs will taper the waveform as it propagates throughout the crystal and create an effective optical iris with quadratic attenuation. Though this effect is negligible under typical operation, during high chirp rate operations that require fast fade-in and fade-out tones with non-negligible $A^{(2)}(t)$ terms, acoustic irising will dilate tweezer waist sizes.



\bibliography{library}


\clearpage

\end{document}